\documentclass[journal,draftclsnofoot,onecolumn,12pt]{IEEEtran}
\ifCLASSINFOpdf
\else
\fi

\usepackage{graphicx}
\usepackage{bm}
\usepackage{amsfonts}
\usepackage{amsmath}
\usepackage{amssymb}
\usepackage{times}
\usepackage{subfigure}
\usepackage{latexsym,bm,amsmath,amssymb} 
\usepackage{cite}
\usepackage{hhline}

\usepackage{multirow} 
\usepackage{amsmath}
\usepackage{xcolor}
\usepackage{epstopdf}
\usepackage{algorithmicx,algorithm}

\usepackage{geometry}
\begin{document}

%
\title{Delay-Doppler Domain Tomlinson-Harashima Precoding for OTFS-based Downlink MU-MIMO Transmissions: Linear Complexity Implementation and Scaling Law Analysis}

\author{Shuangyang~Li,~\IEEEmembership{Member,~IEEE,}
Jinhong~Yuan,~\IEEEmembership{Fellow,~IEEE,}
Paul~Fitzpatrick,~\IEEEmembership{Senior Member,~IEEE,}
Taka~Sakurai,~\IEEEmembership{Member,~IEEE,} and
Giuseppe~Caire,~\IEEEmembership{Fellow,~IEEE}
\thanks{
   Part of the paper was presented at IEEE Global Communications Conference 2022~\cite{MIMO_conference_THP}.

}
}


\maketitle
\vspace{-10mm}
\begin{abstract}
Orthogonal time frequency space (OTFS) modulation is a recently proposed delay-Doppler (DD) domain communication scheme, which has shown promising performance in general wireless communications, especially over  high-mobility channels.
In this paper, we investigate DD domain Tomlinson-Harashima precoding (THP) for downlink multiuser multiple-input and multiple-output OTFS (MU-MIMO-OTFS) transmissions. Instead of directly applying THP based on the huge equivalent channel matrix, we propose a simple implementation of THP that does not require any matrix decomposition or inversion. Such a simple implementation is enabled by the DD domain channel property, i.e., different resolvable paths do not share the same delay and Doppler shifts, which makes it possible to pre-cancel all the DD domain interference in a symbol-by-symbol manner. We also study the achievable rate performance for the proposed scheme by leveraging the information-theoretical equivalent models. In particular, we show that the proposed scheme can achieve a near optimal performance in the high signal-to-noise ratio (SNR) regime. More importantly, scaling laws for achievable rates with respect to number of antennas and users are derived, which indicate that the achievable rate increases logarithmically with the number of antennas and linearly with the number of users. Our numerical results align well with our findings and also demonstrate a significant improvement compared to existing MU-MIMO schemes on OTFS and orthogonal frequency-division multiplexing (OFDM).
\end{abstract}


\IEEEpeerreviewmaketitle
\section{Introduction}
Orthogonal time frequency space (OTFS) modulation has received much attention in the past few years since its invention in~\cite{Hadani2017orthogonal}, thanks to its capability of providing highly reliable communications over complex transmission scenarios, such as high-mobility channels~\cite{Zhiqiang_magzine,Shuangyang2021tutorial}. Compared to the currently deployed orthogonal frequency-division multiplexing (OFDM) modulation, OTFS modulation has demonstrated high-Doppler resilience and robust communication performance against various channel conditions~\cite{Zhiqiang_magzine,Hadani2018OTFS_long,gaudio2021otfs,raviteja2019otfs}. Therefore, OTFS modulation has been recognized as a potential solution to supporting the heterogeneous requirements of beyond fifth-generation (B5G) wireless systems, especially in high-mobility scenarios~\cite{Zhiqiang_magzine,Hadani2018OTFS_long}.

The success of OTFS originates from the delay-Doppler (DD) domain signal processing~\cite{lampel2021orthogonal,mohammed2021derivation}, guided by the elegant mathematical theory of the Zak transform~\cite{janssen1988zak,bolcskei1997discrete}. The Zak transform gives rise to the DD domain symbol placement, which potentially enables pulse localization without violating Heisenberg's uncertainty principle~\cite{Hadani2017orthogonal,Hadani2018OTFS_long}. Furthermore, the DD domain symbol placement allows the information symbols to directly interact with the DD domain channel response, resulting in a much simpler input-output relationship compared to that of OFDM modulation over complex channels such as the high-mobility channel. More importantly, it can be shown that with DD domain modulation, each information symbol principally experiences the whole fluctuations of the time-frequency (TF) channel over an OTFS frame. Thus, the OTFS modulation offers the potential of achieving full TF diversity~\cite{Surabhi2019on,Raviteja2019effective,li2020performance,Chong2022achievable,Ruoxi2022outage}.

The DD domain channel response has several appealing properties including compactness, quasi-stationarity, separability, and sparsity~\cite{hlawatsch2011wireless,Herbert2019sparsity}, which enables simple channel estimation and reduced-complexity detection approaches.
For example, an embedded pilot scheme for OTFS channel estimation
was proposed in~\cite{raviteja2019embedded}, where a sufficiently large guard interval is applied around the pilot to improve the acquisition of delay and Doppler responses. Such a scheme can permit a direct channel estimation by simply checking the received signal's value around the DD grid of the embedded pilot.
In~\cite{Wei2022off}, a sparse Bayesian-learning-assisted channel estimation approach was presented, where both on-grid and off-grid (due to the virtual sampling) delay and Doppler components are used to perform sparse signal recovery in order to estimate the delay and Doppler responses.
A message passing algorithm (MPA) was proposed in~\cite{Raviteja2018interference}, where the Gaussian approximation is applied to model the characteristic of DD domain interference. This algorithm and its variants, such as~\cite{Yuan2020simple},~\cite{li2021hybrid}, and~\cite{ZhengdaoYuan2022TWC}, take advantage of the DD domain sparsity, such that fewer iterations over the graphical model are sufficient to obtain a good error performance.
The aforementioned algorithms and many other excellent works~\cite{Chockalingam2020LowComplexity,TharajRake} have laid a strong foundation for single-input and single-output (SISO)-OTFS transceiver designs. However, related investigations on multiple-input and multiple-output (MIMO)-OTFS systems are only in the their infancy.

MIMO technology is an important candidate to meet the stringent requirements of the achievable rate for B5G wireless systems~\cite{tse2005fundamentals}. Research on MIMO-OTFS, especially multiuser MIMO-OTFS (MU-MIMO-OTFS), is important to determine whether OTFS modulation can be applied in practical multiple-antenna systems~\cite{mohammadi2022cell}. Unfortunately, the design of MU-MIMO-OTFS is challenging. This is because OTFS modulation does not guarantee interference-free transmission like OFDM modulation in static channels. In fact, the DD domain received symbols generally contain interference~\cite{Raviteja2018interference} in the multi-path transmission, as the result of the ``twisted convolution'' between the transmitted symbols and the DD domain channel responses{\footnote{The term ``twisted convolution'' comes from the first OTFS paper~\cite{Hadani2017orthogonal}, which is similar to the circular convolution but
with an additional phase term.}}~\cite{Hadani2018OTFS_long}.  Consequently, most of the designs of MU-MIMO-OTFS will face an equivalent channel matrix with a huge size, e.g., number of delay bins times number of Doppler bins times number of antennas. With such an enormous matrix size, conventional precoding/equalization techniques, such as zero forcing and minimum mean square error (MMSE), cannot be directly applied due to the extremely high computational complexity introduced by the channel inversion. As a result, most of the existing works for downlink MU-MIMO-OTFS rely on simple precoding approaches, such as maximum ratio transmission (MRT) precoding~\cite{Pandey2021low}, or approximation of channel inversion, such as~\cite{cao2021low}, with an aim to reduce the computational complexity by trading off performance.

In this paper, we consider the precoding design for downlink MU-MIMO-OTFS from a different perspective by using the Tomlinson-Harashima precoding (THP)~\cite{tomlinson1971new,Harashima1972matched}.
THP is a classic non-linear precoding scheme that has been widely applied in practice, whose core idea is to pre-cancel/pre-subtract the known interference before transmission.
THP has shown promising performance in terms of the achievable rate. In particular, it has been shown in~\cite{Wesel1998Achievable} that the constant ``shaping loss'' is the only loss of the achievable rate for THP at high signal-to-noise ratios (SNRs)~\cite{Wesel1998Achievable}. Thus, we postulate that the application of THP in MU-MIMO-OTFS would result in a promising rate performance. Note that the conventional implementation of THP requires QR decomposition~\cite{tomlinson1971new,Harashima1972matched}, such that the decomposed channel matrix has a triangular structure. However, with a huge matrix size in the MU-MIMO-OTFS transmission, such a decomposition could be computationally expensive.
In contrast to the existing works, we do not aim to design precoding directly based on the huge equivalent channel matrix. Instead, we propose to perform interference pre-cancellation directly in the DD domain without any channel decomposition or inversion. This is possible by exploiting the fact that different resolvable paths must be distinguishable in at least one dimension of delay and Doppler, and consequently cannot share both the same delay and Doppler shifts at the same time{\footnote{Physical channels can have multiple paths sharing the same or very similar delay and Doppler responses. However, due to the limited capability of distinguishing delay and Doppler for practical receivers, those paths cannot be fully resolved or separated. Consequently, the receiver only sees one multi-path component (DD response) due to the combining of these paths~\cite{sayeed2010wireless}.}~\cite{hlawatsch2011wireless,sayeed2010wireless}.
The major contributions of this paper can be summarized as follows.
\begin{itemize}
\item We derive a concise input-output relation for downlink MU-MIMO-OTFS with beamforming (BF) in the matrix form, which lays the foundations for our digital precoder designs and later performance analysis.
\item Using the derived system model, we conduct a detailed analysis on the DD domain interference pattern and compare it to the TF domain interference pattern for the OFDM counterpart. In particular, we show that the DD domain received symbols suffer from three types of interference, namely multi-path self-interference (MPSI), inter-beam interference (IBI), and crosstalk interference (CTI). We unveil the physical meanings of those interference terms, and show that IBI can be ignored by considering user grouping or user scheduling, while MPSI can be mitigated by BF in practical systems.
\item We propose a DD domain THP design that only entails linear complexity without any matrix decomposition or inversion based on the characteristics of DD domain channel responses. In particular, we show that the DD domain interference pattern contains several cycles. The existence of the cycles suggests that the interference pre-cancellation can start from any DD grid in the cycle and all the interference can be cancelled out in a symbol-by-symbol manner.
\item We study the sum-rate of the proposed scheme by deriving the representative information-theoretical equivalent models according to the property of the modulo operation. Based on the derived sum-rate, we show that the proposed scheme can achieve a near-optimal performance that only has a constant rate loss (the shaping loss) compared to the optimal interference-free transmission. Furthermore, we investigate the sum-rate performance with respect to the number of antennas at the base station (BS) $N_{\rm BS}$ and the number of users $K$, respectively. In particular, we show that the sum-rate of the proposed scheme increases linearly with $K$ and logarithmically with $N_{\rm BS}$.
\end{itemize}


\emph{Notations:} The blackboard bold letters ${\mathbb{A}}$, ${\mathbb{E}}$, and ${\mathbb{C}}$ denote the constellation set, the expectation operator, and the complex number field, respectively; the notations $(\cdot)^{\rm{T}}$ and $(\cdot)^{\rm{H}}$ denote the transpose and the Hermitian transpose for a matrix, respectively; $\textrm{vec}(\cdot)$ denotes the vectorization operation; $\textrm{diag}{\{\cdot\}}$ denotes the diagonal matrix; ``$ \otimes $" denotes the Kronecker product operator; $\min \left( {\cdot} \right)$ returns the minimum value of a function; $I\left( \cdot ; \cdot \right)$ and $h\left( \cdot \right)$ denote the mutual information and the differential entropy, respectively;
${\left[ {\cdot} \right]_x}$ denotes the modulo operation with respect to $x$. ${{{\bf{F}}_N}}$ and ${{{\bf{I}}_M}}$ denote the discrete Fourier transform (DFT) matrix of size $N\times N$ and the identity matrix of size $M\times M$; the big-O notation ${\cal O}\left( \cdot \right)$ describes the asymptotic growth rate of a function. For the sake of clarity, the main system parameters are summarized in Table~\ref{notations_table}.

\begin{table}[]
\caption{List of Main System Parameters.}
\small
\centering
\begin{tabular}{|c|c|}
\hline
Parameters&Definitions\\ \hline
$K$ & Number of users  \\ \hline
$M$ & Number of delay bins/subcarriers   \\ \hline
$N$ & Number of Doppler bins/time slots  \\ \hline
$P$ & Number of resolvable paths  \\ \hline
$N_{\rm BS}$& Number of antennas at BS \\ \hline
$\Delta f$ & Subcarrier spacing   \\ \hline
$T$ & Time slot duration  \\ \hline
$L$ & Number of interference terms considered for cancellation \\ \hline
$h_p^{\left( k \right)}$& Channel coefficient for the $p$-th path of the $k$-th user \\ \hline
$l_p^{\left( k \right)}$ and $k_p^{\left( k \right)}$ & Delay and Doppler indices for the $p$-th path of the $k$-th user \\ \hline
${g_p^{\left( i \right)}\left[ j \right]}$ & Spatial interference power of the $j$-th beam on $i$-th user's $p$-th path \\ \hline
$X_{{\rm{DD}}}^{\left( i \right)}\left[ {l,k} \right]$ and $Y_{{\rm{DD}}}^{\left( i \right)}\left[ {l,k} \right]$ & $(l,k)$-th DD domain transmitted and received symbol of the $i$-th user\\ \hline
\end{tabular}
\label{notations_table}
\end{table}
\section{System Model}
In this section, we will derive a concise system model for MU-MIMO-OTFS transmissions. Before going into the details of MU-MIMO-OTFS transmissions, we will briefly review some preliminaries on SISO-OTFS transmissions, which will then be used for the related discussions on MU-MIMO-OTFS transmissions.
\subsection{Preliminaries on SISO-OTFS Transmissions}
\begin{figure}
\centering
\includegraphics[width=0.8\textwidth]{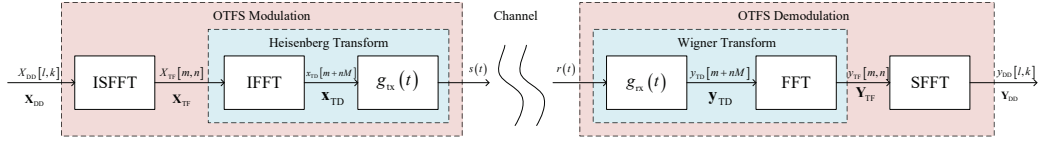}
\caption{The transmitter structure of SISO-OTFS transmissions.}
\label{System_Model_SISO}
\centering
\end{figure}
Without loss of generality, let us consider the OTFS transmitter shown in Fig.~\ref{System_Model_SISO}. Let $M$ be the number of delay bins/subcarriers and $N$ be the number of Doppler bins/time slots, respectively. The corresponding subcarrier spacing and time slot duration are given by $\Delta f$ and $T$, respectively. Let ${\bf{x}}_{\rm DD} \in {{\mathbb{A}}^{M N}}$ be the DD domain information symbol vector of length $MN$.
In particular, the information symbol vector ${\bf{x}}_{\rm DD}$ can be arranged as a two-dimensional (2D) information symbol matrix ${\bf{X}}_{\rm DD} \in {{\mathbb{A}}^{M \times N}}$, i.e., ${\bf{x}}_{\rm DD} \buildrel \Delta \over = \textrm{vec}\left( {\bf{X}}_{\rm DD} \right)$, and the $(l,k)$-th element of ${\bf{X}}_{\rm DD}$, $X_{\rm DD}\left[ {l,k} \right]$, is the information symbol at the $l$-th delay grid and the $k$-th Doppler grid~\cite{Hadani2017orthogonal}, for $0 \le k \le N-1,0 \le l \le M-1$.
As indicated by Fig.~\ref{System_Model_SISO}, the TF domain transmitted symbol $X_{\rm TF}\left[ {m,n} \right], 0 \le m \le M-1, 0 \le n \le N-1$ can be obtained from ${\bf{X}}_{\rm DD}$ via the inverse symplectic finite Fourier transform (ISFFT)~\cite{Raviteja2019practical}, i.e.,
\begin{equation}
{{\bf{X}}_{{\rm{TF}}}} \buildrel \Delta \over = {{\bf{F}}_M}{{\bf{X}}_{{\rm{DD}}}}{\bf{F}}_N^{\rm{H}}  , \label{SISO_ISFFT}
\end{equation}
where $X_{\rm TF}\left[ {m,n} \right]$ is the $\left( {m,n} \right)$-th element in ${\bf{X}}_{\rm TF}$, and ${\bf F}_M$ and ${\bf F}_N$ are the normalized DFT matrices of size $M \times M$ and $N \times N$ defined in the \emph{Notations}. It is also convenient to write the corresponding vector form of~\eqref{SISO_ISFFT}, which is given by~\cite{li2021cross}
\begin{equation}
{{\bf{x}}_{{\rm{TF}}}} \buildrel \Delta \over = {\rm{vec}}\left( {{{\bf{X}}_{{\rm{TF}}}}} \right) = \left( {{\bf{F}}_N^{\rm{H}} \otimes {{\bf{F}}_M}} \right){{\bf{x}}_{{\rm{DD}}}} . \label{SISO_ISFFT_vec}
\end{equation}
The transmitted OTFS signal $s\left( t \right)$ can be obtained by performing the Heisenberg transform~\cite{Hadani2017orthogonal} to ${\bf{X}}_{\rm TF}$ with the transmitter shaping pulse $g_{{\rm{tx}}}(t)$, as shown in Fig.~\ref{System_Model_SISO}.
In particular, the Heisenberg transform can be interpreted as a multicarrier modulator and a popular choice for implementing the Heisenberg transform is to apply the OFDM modulator~\cite{Zhiqiang_magzine}.
According to the OFDM modulation, the Heisenberg transform can be implemented by an inverse fast Fourier transform (IFFT) module and transmit pulse shaping, in which case the resultant transmitted OTFS signal $s\left( t \right)$ is given by
\begin{equation}
s\left( t \right) = \sum\limits_{n = 0}^{N - 1} {\sum\limits_{m = 0}^{M - 1} {X_{\rm TF}\left[ {m,n} \right]{g_{{\rm{tx}}}}\left( {t - nT} \right){e^{j2\pi m\Delta f\left( {t - nT} \right)}}} }.\label{SISO_OTFS_signal}
\end{equation}
Based on~\eqref{SISO_OTFS_signal}, it is useful to define the time-delay (TD) domain transmitted symbol vector ${\bf{x}}_{\rm TD}$ of length $MN$. Considering the energy-normalized rectangular shaping pulse ${g_{{\rm{tx}}}}\left( t \right)$, ${\bf{x}}_{\rm TD}$ is defined by~\cite{Raviteja2019practical}
\begin{equation}
{{\bf{x}}_{{\rm{TD}}}} \buildrel \Delta \over = {\rm{vec}}\left( {{\bf{F}}_M^{\rm{H}}{{\bf{X}}_{{\rm{TF}}}}} \right) = \left( {{\bf{F}}_N^{\rm{H}} \otimes {{\bf{I}}_M}} \right){{\bf{x}}_{{\rm{DD}}}}.
\label{SISO_OTFS_TD_symbol_vec}
\end{equation}

Let ${h_{{\rm{DD}}}}\left( {\tau ,\nu } \right)$ be the DD domain channel response given by
\begin{align}
h_{{\rm{DD}}}\left( {\tau ,\nu } \right) = \sum\limits_{p = 1}^P {{h_p}} \delta \left( {\tau  - {\tau _p}} \right)\delta \left( {\nu  - {\nu _p}} \right),   \label{SISO_channel_DD}
\end{align}
where ${h_p}$, ${\tau _p}$, and ${\nu _p}$ are the fading coefficient, the delay shift, and the Doppler shift associated with the $p$-th path.

According to~\cite{Raviteja2019practical}, the corresponding TD domain channel response of~\eqref{SISO_channel_DD} can be equivalently represented in a matrix form in the case of rectangular filtering pulse ${g_{{\rm{rx}}}}\left( t \right)$, reduced CP structure, and non-fractional delay and Doppler shifts, such that
\begin{equation}
{\bf{H}}_{\rm{TD}} = \sum\limits_{p = 1}^P {{h_p}} {{\bm{\Pi }}^{{l_p}}}{{\bm{\Delta}} ^{{k_p}}}, \label{SISO_TD_channel}
\end{equation}
where ${\bm{\Pi }}$ is the permutation matrix (forward cyclic shift), i.e.,
\begin{equation}
{\bm{\Pi }} = \small {\left[ {\begin{array}{*{20}{c}}
0& \cdots &0&1\\
1& \ddots &0&0\\
 \vdots & \ddots & \ddots & \vdots \\
0& \cdots &1&0
\end{array}} \right]_{MN \times MN}},
\end{equation}
and ${\bm{\Delta}}=\textrm{diag}\{{\gamma}^0,{\gamma}^1,...,{\gamma}^{MN-1}\} $ is a diagonal matrix with ${\gamma} \buildrel \Delta \over = {e^{\frac{{j2\pi }}{{MN}}}}$~\cite{Raviteja2019practical}.
In~\eqref{SISO_TD_channel}, the terms $l_p$ and $k_p$ are the indices of delay and Doppler, respectively, associated with the $p$-th path, respectively, where
\begin{equation}
{\tau _p} = \frac{{l_p}}{{M\Delta f}},\quad {\rm and }\quad
{\nu _p} = \frac{{{k_p}}}{{NT}},
\label{resolution}
\end{equation}
and we have ${l_p} \le {l_{\max }}$ and $- {k_{\max }} \le {k_p} \le {k_{\max }}$, for $1 \le p \le P$, with $l_{\rm max}$ and $k_{\rm max}$ denoting the largest delay index and Doppler index, respectively.
It should be noted that the system model in~\eqref{SISO_TD_channel} only considers the integer delay and Doppler case, which is only valid with a sufficiently large signal bandwidth and a sufficiently long frame duration~\cite{Raviteja2018interference}. However, it is reported in~\cite{wei2021transmitter} that the effects of fractional Doppler could be mitigated by adding TF domain windows. Furthermore, some recent developments of OTFS have shown that the pulse shaping could improve the DD domain sparsity~\cite{mohammed2021derivation,lampel2021orthogonal,Hai2022TWC,Wei2022CL,Shuangyang2022CL}.
As the main focus of this paper is on the application of THP to MU-MIMO-OTFS transmissions, we restrict ourselves to the case of integer delay and Doppler.
Following on from~\eqref{SISO_TD_channel}, the received time-delay (TD) domain symbol vector ${\bf{y}}_{\rm TD}$ is given by
\begin{equation}
{\bf{y}}_{\rm TD} = {\bf{H}}_{\rm{TD}}{{\bf{x}}_{{\rm{TD}}}} + {\bf{w}} \label{SISO_TD_io_relationship},
\end{equation}
where ${\bf{w}}$ is the corresponding additive white Gaussian noise (AWGN) sample vector in the TD domain with one-sided power spectral density (PSD) $N_0$.
The OTFS demodulation can be interpreted as the concatenation of the Wigner transform and the SFFT~\cite{Hadani2017orthogonal}. Based on~\eqref{SISO_TD_io_relationship}, the DD domain received symbol vector is given by{\footnote{In~\eqref{SISO_DD_io_relationship}, we use the same notation for the AWGN samples in both TD and DD domains, because they follow the same distribution.}}~\cite{Raviteja2019practical},
\begin{align}
{{\bf{y}}_{{\rm{DD}}}} = \left( {{{\bf{F}}_N} \otimes {{\bf{I}}_M}} \right){{\bf{y}}_{{\rm{TD}}}}={\bf{H}}_{\rm{DD}}{{\bf{x}}_{{\rm{DD}}}} + {\bf{w}} \label{SISO_DD_io_relationship},
\end{align}
where ${\bf{H}}_{\rm{DD}}$ is the corresponding equivalent DD domain channel matrix of the form~\cite{li2020performance}
\begin{align}
{\bf{H}}_{\rm{DD}} \buildrel \Delta \over = \sum\limits_{p = 1}^P {{h_p}\left( {{{\bf{F}}_N} \otimes {{\bf{I}}_M}} \right)} {{\bf{\Pi }}^{{l_p}}}{{\bf{\Delta }}^{{k_p} }}\left( {{\bf{F}}_N^{\rm{H}} \otimes {{\bf{I}}_M}} \right).
\label{SISO_DD_channel}
\end{align}

For ease of derivation, it is useful to derive a DD domain symbol-wise input-output relation based on~\eqref{SISO_DD_io_relationship}. In fact,~\eqref{SISO_DD_channel} has a direct connection to the inverse discrete Zak transform (IDZT), which gives rise to the following lemma.

\textbf{Lemma 1} (\emph{DD Domain Input-Output Relation via IDZT}):
Let ${\bf Y}_{\rm DD}$ be the corresponding matrix representation of ${\bf y}_{\rm DD}$, i.e., ${{\bf{y}}_{{\rm{DD}}}} \buildrel \Delta \over = {\rm{vec}}\left( {{{\bf{Y}}_{{\rm{DD}}}}} \right)$. Then, in the case of integer Doppler indices and rectangular shaping pulses, the input-output relation for OTFS transmissions with the reduced-CP structure and without noise can be characterized by
\begin{align}
Y_{\rm DD}\left[ {l,k} \right] = \sum\limits_{p = 1}^P {{h_p}{e^{j2\pi \frac{{k_p\left( {l - {l_p}} \right)}}{{MN}}}}{\alpha _{l,l_p,k,k_p}}X_{\rm DD}\left[ {{{\left[ {l - {l_p}} \right]}_M},{{\left[ {k - {k_p}} \right]}_N}} \right]}  , \label{SISO_IO_DZT}
\end{align}
where ${\alpha _{l,l_p,k,k_p}}$ is a phase offset as the result of the quasi-periodicity property of the IDZT, and it is given by
\begin{align}
{\alpha _{l,l_p,k,k_p}} = \left\{ \begin{array}{l}
1,\quad\quad\quad\quad \quad\quad  l - {l_p} \ge 0,\\
{e^{ - j2\pi \frac{{k - {k_p}}}{N}}},\quad\quad\quad\!\! l - {l_p} < 0.
\end{array} \right. \label{SISO_IO_phase_offset}
\end{align}

\textbf{Proof}: The proof is straightforward by invoking the IDZT. Furthermore, derivations without applying IDZT can also be found in Section 4.6.2 of~\cite{Viterbo2022DDcommunications}.

Despite the fact that Lemma~1 has already appeared in the literature~\cite{Viterbo2022DDcommunications}, we still want to emphasize the importance of those results here because of the following two reasons. Firstly, the symbol-wise DD domain input-output relation for OTFS has not been widely considered and understood in the literature. Secondly, the results of Lemma~1 will be frequently used in the later part of this paper as the building block for our derivations.
Based on the above descriptions of SISO-OTFS transmissions, we will der ive the system model of MU-MIMO-OTFS transmissions in the following subsection.
\subsection{Derivations of the System Model for MU-MIMO-OTFS Transmissions}
\begin{figure}
\centering
\includegraphics[width=0.8\textwidth]{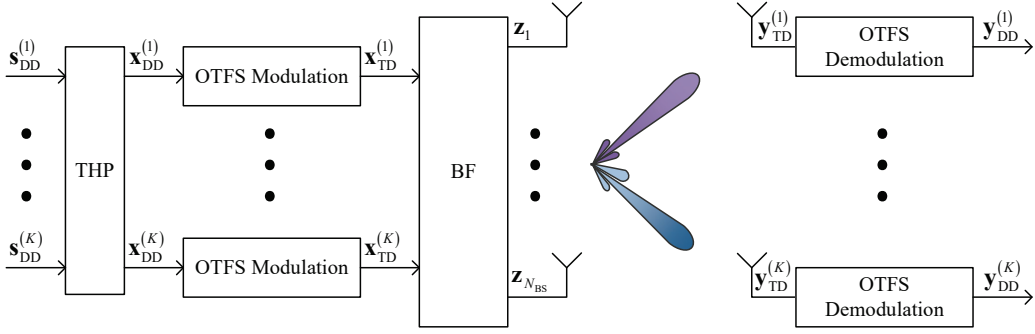}
\caption{The block diagram of considered THP-based downlink MU-MIMO-OTFS transmissions.}
\label{System_Model_MU_MIMO}
\centering
\end{figure}
Without loss of generality, let us consider the downlink MU-MIMO-OTFS transmission for $K$ users, where the BS is equipped with $K$ radio-frequency (RF) chains and $N_{\rm BS}$ antennas with $N_{\rm BS} \ge K$, while each user is equipped with only one antenna, as shown in Fig.~\ref{System_Model_MU_MIMO}. For notational consistency, we will extend the related notations from the above subsection by adding superscripts or subscripts to specify the underlying users or antennas. Denote by ${\bf{s}}_{{\rm{DD}}}^{\left( k \right)} \in {{\mathbb A}^{MN \times 1}}$ the DD domain information symbol vector of length $MN$ for the $k$-th user, where $1 \le k \le K$. In particular, the DD domain information symbol vectors for the $K$ users can be arranged into a 2D matrix ${\bf S}_{\rm DD}$ of size $MN \times K$, whose $k$-th column is ${\bf{s}}_{{\rm{DD}}}^{\left( k \right)}$.
As indicated by Fig.~\ref{System_Model_MU_MIMO}, we apply THP to ${\bf S}_{\rm DD}$ and the resultant symbol matrix after precoding is ${\bf X}_{\rm DD}$ of size $MN \times K$, whose $k$-th column is the DD domain symbol vector for the $k$-th user after precoding, denoted by ${\bf{x}}_{{\rm{DD}}}^{\left( k \right)}$.
After passing ${\bf{x}}_{{\rm{DD}}}^{\left( k \right)}$ through the OTFS modulator, the TD domain symbol vector for the $k$-th user can be obtained by ${\bf{x}}_{{\rm{TD}}}^{\left( k \right)}=\left( {{\bf{F}}_N^{\rm{H}} \otimes {{\bf{I}}_M}} \right){\bf{x}}_{{\rm{DD}}}^{\left( k \right)}$ according to~\eqref{SISO_OTFS_TD_symbol_vec}.
Thus, we can write
\begin{equation}
{{\bf{X}}_{{\rm{TD}}}} = \left( {{\bf{F}}_N^{\rm{H}} \otimes {{\bf{I}}_M}} \right){{\bf{X}}_{{\rm{DD}}}}, \label{MIMO_OTFS_TD_symbol_matrix}
\end{equation}
where ${{\bf{X}}_{{\rm{TD}}}}$ of size $MN \times K$ is the TD domain symbol matrix after OTFS modulation, and its $k$-th column is ${\bf{x}}_{{\rm{TD}}}^{\left( k \right)}$.
For ease of derivation, let us consider the vectorized version of ${{\bf{X}}_{{\rm{TD}}}}$ by stacking each column of ${{\bf{X}}_{{\rm{TD}}}}$ into a vector, such as
\begin{align}
{{\bf{x}}_{{\rm{TD}}}} \buildrel \Delta \over = \left[ {{{\left( {{\bf{x}}_{{\rm{TD}}}^{\left( 1 \right)}} \right)}^{\rm{H}}},{{\left( {{\bf{x}}_{{\rm{TD}}}^{\left( 2 \right)}} \right)}^{\rm{H}}},...,{{\left( {{\bf{x}}_{{\rm{TD}}}^{\left( K \right)}} \right)}^{\rm{H}}}} \right]^{\rm H} = {\rm{vec}}\left( {{{\bf{X}}_{{\rm{TD}}}}} \right) = \left( {{{\bf{I}}_K} \otimes {\bf{F}}_N^{\rm{H}} \otimes {{\bf{I}}_M}} \right){{\bf{x}}_{{\rm{DD}}}}, \label{TD_MU_MIMO_vec}
\end{align}
where ${{\bf{x}}_{{\rm{DD}}}} \buildrel \Delta \over = {\rm{vec}}\left( {{{\bf{X}}_{{\rm{DD}}}}} \right)$ is the DD domain symbol vector of size $KMN \times 1$.
We consider conventional BF for the downlink transmission as indicated in Fig.~\ref{System_Model_MU_MIMO}. Let ${{\bf{V}}_{{\rm{BF}}}}$ of size $K \times N_{\rm BS}$ be the BF matrix adopted. Then, the transmitted symbol matrix ${\bf{Z}}$ after BF is given by
\begin{align}
{\bf{Z}} = {{\bf{X}}_{{\rm{TD}}}}{{\bf{V}}_{{\rm{BF}}}}, \label{Z_BF_MU_MIMO_matrix}
\end{align}
where the $n$-th column of ${\bf{Z}}$, ${{\bf z}_{{n}}}$, is the transmitted symbol vector on the $n$-th antenna at the BS, for $1\le n \le N_{\rm BS}$. Similar to~\eqref{TD_MU_MIMO_vec}, we can write the corresponding vector form of~\eqref{Z_BF_MU_MIMO_matrix}, which is given by
\begin{align}
{\bf{z}} \buildrel \Delta \over = \left[ {{\bf{z}}_1^{\rm{H}},{\bf{z}}_2^{\rm{H}},...,{\bf{z}}_{{N_{{\rm{BS}}}}}^{\rm{H}}} \right]^{\rm H} = {\rm{vec}}\left( {\bf{Z}} \right) = \left( {{\bf{V}}_{{\rm{BF}}}^{\rm{T}} \otimes {{\bf{I}}_{MN}}} \right){{\bf{x}}_{{\rm{TD}}}} = \left( {{\bf{V}}_{{\rm{BF}}}^{\rm{T}} \otimes {\bf{F}}_N^{\rm{H}} \otimes {{\bf{I}}_M}} \right){{\bf{x}}_{{\rm{DD}}}}.\label{Z_BF_MU_MIMO_vec}
\end{align}

Now let us turn our attention to the wireless channel for MU-MIMO transmissions.
Without loss of generality, we assume that the antenna array at the BS is in the form of
a uniform linear array (ULA).
We further assume that the underlying channel between the BS and each user has
$P$ independent resolvable paths, where the angle-of-departure (AoD) for the $p$-th path of the $k$-th user, for $1 \le p \le P$ and $1 \le k \le K$, is given by $\varphi _p^{\left( k \right)}$, and $\varphi _p^{\left( k \right)} \ne \varphi _{p'}^{\left( {k'} \right)}$, for $p \ne p'$ or $k \ne k'$. Then, according to the far field assumption~\cite{tse2005fundamentals} and the DD domain channel characteristics in~\eqref{SISO_channel_DD}, the DD domain channel for the $n$-th antenna and the $k$-th user can be modeled by
\begin{align}
h\left( {n,k,\tau ,\nu } \right)
= \sum\limits_{p = 1}^P {h_p^{\left( k \right)}\exp \left( {j\pi \left( {n - 1} \right)\sin \left(\varphi _p^{\left( k \right)}\right)} \right)} \delta \left( {\tau  - \tau _p^{\left( k \right)}} \right)\delta \left( {\nu  - \nu _p^{\left( k \right)}} \right),   \label{MU_MIMO_channel_DD1}
\end{align}
where we assume that the distance between adjacent antennas is equal to half of the wavelength. In~\eqref{MU_MIMO_channel_DD1}, $h_p^{\left( k \right)}\in {\mathbb{C}}$, $\tau _p^{\left( k \right)}$, and $\nu _p^{\left( k \right)}$ are the fading coefficient, the delay shift, and the Doppler shift corresponding to the $p$-th path of the $k$-th user, respectively.
According to~\eqref{MU_MIMO_channel_DD1}, let us denote by $l_p^{\left( k \right)}$ and $k_p^{\left( k \right)}$ the delay and Doppler indices corresponding to the $p$-th path of the $k$-th user, i.e.,
\begin{equation}
\tau _p^{\left( k \right)} = \frac{{{l_p^{\left( k \right)}}}}{{M\Delta f}},\quad
\nu _p^{\left( k \right)} = \frac{{{k_p^{\left( k \right)}}}}{{NT}}.
\label{MU_MIMO_channel_resolution}
\end{equation}
Let us further define the  effective TD domain channel matrix for the $p$-th path of the $k$-th user based on~\eqref{SISO_TD_channel} by ${\bf{\tilde H}}_{{\rm{TD}}}^{k,p} = {h_p^{\left( k \right)}} {{\bf{\Pi }}^{l_p^{\left( k \right)}}}{{\bf{\Delta }}^{k_p^{\left( k \right)}}}$.
Similarly, based on~\eqref{SISO_DD_channel}, the effective DD domain channel matrix for the $p$-th path of the $k$-th user is defined by ${\bf{\tilde H}}_{{\rm{DD}}}^{k,p} \buildrel \Delta \over =  {h_p^{\left( k \right)}\left( {{{\bf{F}}_N} \otimes {{\bf{I}}_M}} \right)} {{\bf{\Pi }}^{l_p^{\left( k \right)}}}{{\bf{\Delta }}^{k_p^{\left( k \right)}}}\left( {{\bf{F}}_N^{\rm{H}} \otimes {{\bf{I}}_M}} \right)$.
After some derivations, we can write the TD domain received
symbol vector ${{\bf{y}}_{\rm TD}^{\left( k \right)}}$ for the $k$-th user by
\begin{align}
{{\bf{y}}_{\rm TD}^{\left( k \right)}}
= {\sqrt {{N_{{\rm{BS}}}}} }\sum\limits_{p = 1}^P {\left( { {{\bf{a}}^{\rm{T}}}\left( {{\varphi _p^{\left( k \right)}}} \right) \otimes {\bf{\tilde H}}_{\rm{TD}}^{k,p}} \right){\bf{z}} + {\bf{w}}^{\left( k \right)}}, \label{MU_TD_IO_relationship3}
\end{align}
where
\begin{align}
{\bf{a}}\left( {\varphi _p^{\left( k \right)}} \right) \buildrel \Delta \over = \frac{1}{{\sqrt {{N_{{\rm{BS}}}}} }}{\left[ {1,\exp \left( {j\pi \sin \varphi _p^{\left( k \right)}} \right),...,\exp \left( {j\pi \left( {{N_{{\rm{BS}}}} - 1} \right)\sin \varphi _p^{\left( k \right)}} \right)} \right]^{\rm{T}}}, \label{MU_steering_vector}
\end{align}
is the normalized steering vector for the $p$-th path of the $k$-th user, and ${\bf{w}}^{\left( k \right)}$ is the AWGN sample vector with one-sided PSD $N_0$.
Next, by considering~\eqref{Z_BF_MU_MIMO_vec},~\eqref{MU_TD_IO_relationship3} can be further expanded as
\begin{align}
{\bf{y}}_{{\rm{TD}}}^{\left( k \right)} &= {\sqrt {{N_{{\rm{BS}}}}} }\sum\limits_{p = 1}^P {\left( {\left( {{{\bf{a}}^{\rm{T}}}\left( {\varphi _p^{\left( k \right)}} \right){\bf{V}}_{{\rm{BF}}}^{\rm{T}}} \right) \otimes \left( {{\bf{\tilde H}}_{{\rm{TD}}}^{k,p}\left( {{\bf{F}}_N^{\rm{H}} \otimes {{\bf{I}}_M}} \right)} \right)} \right)} {{\bf{x}}_{{\rm{DD}}}} + {{\bf{w}}^{\left( k \right)}}\notag\\
 &= {\sqrt {{N_{{\rm{BS}}}}} }\sum\limits_{p = 1}^P {\left( {{\bf{\tilde H}}_{{\rm{TD}}}^{k,p}\left( {{\bf{F}}_N^{\rm{H}} \otimes {{\bf{I}}_M}} \right)} \right)} {{\bf{X}}_{{\rm{DD}}}}\left( {{{\bf{V}}_{{\rm{BF}}}}{\bf{a}}\left( {\varphi _p^{\left( k \right)}} \right)} \right) + {{\bf{w}}^{\left( k \right)}},
\label{receive_T_MU_MIMO_vec}
\end{align}
where the second equation is due to the properties of the Kronecker product. Considering~\eqref{receive_T_MU_MIMO_vec}, it is convenient to define the \emph{effective spatial domain channel vector} ${\bf{g}}_p^{\left( k \right)} \buildrel \Delta \over = {{\bf{V}}_{{\rm{BF}}}}{\bf{a}}\left( {\varphi _p^{\left( k \right)}} \right)$ to characterize the interference from different data streams to the received symbols of the $k$-th user from the $p$-th path.
Finally, by performing OTFS demodulation to ${\bf{y}}_{{\rm{TD}}}^{\left( k \right)}$, the DD domain received symbol vector ${\bf{y}}_{\rm{DD}}^{\left( k \right)}$ for the $k$-th user can be written by
\begin{align}
{\bf{y}}_{{\rm{DD}}}^{\left( k \right)} =& {\sqrt {{N_{{\rm{BS}}}}} }\sum\limits_{p = 1}^P {\left( {\left( {{{\bf{F}}_N} \otimes {{\bf{I}}_M}} \right){\bf{\tilde H}}_{\rm{TD}}^{k,p}\left( {{\bf{F}}_N^{\rm{H}} \otimes {{\bf{I}}_M}} \right)} \right){{\bf{X}}_{{\rm{DD}}}}{\bf{g}}_p^{\left( k \right)}}  + {{\bf{w}}^{\left( k \right)}}\notag\\
 =&{\sqrt {{N_{{\rm{BS}}}}} }\sum\limits_{p = 1}^P {{\bf{\tilde H}}_{{\rm{DD}}}^{k,p}{{\bf{X}}_{{\rm{DD}}}}} {\bf{g}}_p^{\left( k \right)} + {{\bf{w}}^{\left( k \right)}} .\label{receive_DD_MU_MIMO}
\end{align}
So far, we have derived the system model of the MU-MIMO-OTFS transmissions. In the following section, we will develop our digital THP scheme based on~\eqref{receive_DD_MU_MIMO} by adopting a simple BF matrix according to the steering vectors, where the $k$-th row of ${{\bf{V}}_{{\rm{BF}}}}$ is the Hermitian transpose of the steering vector associated with the strongest path of the $k$-th user.

\section{DD Domain THP for Downlink MU-MIMO-OTFS Transmissions}
In this section, we will discuss the proposed DD domain THP. It should be noted that the direct application of THP by employing QR decomposition may require high complexity since the size of the equivalent channel matrix is $KMN \times KMN$.
Therefore, we propose a DD domain THP scheme that does not require the decomposition of channel matrices. In particular, we assume that the channel state information (CSI) is available at the transmitter, which can be achieved by exploiting the DD domain reciprocity~\cite{hlawatsch2011wireless} based on uplink channel estimation.
\subsection{DD Domain Interference Pattern Analysis}
Let us first have a close look at the interference pattern in the DD domain. To provide some insights, let us rewrite~\eqref{receive_DD_MU_MIMO} as
\begin{align}
{\bf{y}}_{{\rm{DD}}}^{\left( i \right)} = {\sqrt {{N_{{\rm{BS}}}}} }\sum\limits_{p = 1}^P {\sum\limits_{j = 1}^K {g_p^{\left( i \right)}\left[ j \right]} } {\bf{\tilde H}}_{{\rm{DD}}}^{i,p}{\bf{x}}_{{\rm{DD}}}^{\left( j \right)} + {{\bf{w}}^{\left( i \right)}}, \label{receive_DD_MU_MIMO_der1}
\end{align}
where ${g_p^{\left( i \right)}\left[ j \right]}$ denotes the $j$-th element of ${\bf{g}}_p^{\left( i \right)} $ implying the contribution from the $j$-th beam to the $i$-th user via the $i$-th user's $p$-th path.
As implied by~\eqref{receive_DD_MU_MIMO_der1}, the DD domain received symbol vector of the $i$-th user is related to the DD domain transmitted symbols of each user.
Furthermore, by considering~\eqref{SISO_IO_DZT},~\eqref{receive_DD_MU_MIMO_der1} can be expanded as
\begin{align}
Y_{{\rm{DD}}}^{\left( i \right)}\left[ {l,k} \right] = \sum\limits_{p = 1}^P {\sum\limits_{j = 1}^K {\tilde g_{l,l_p^{\left( i \right)},k,k_p^{\left( i \right)},p}^{\left( {i,j} \right)}} } X_{{\rm{DD}}}^{\left( j \right)}\left[ {{{\left[ {l - l_p^{\left( i \right)}} \right]}_M},{{\left[ {k - k_p^{\left( i \right)}} \right]}_N}} \right] + {{{w}}^{\left( i \right)}\left[ {l,k} \right]}
, \label{X_Y_i_o_relationship}
\end{align}
where $Y_{{\rm{DD}}}^{\left( i \right)}\left[ {l,k} \right]$ denotes the ${\left( {l,k} \right)}$-th symbol of the received symbol matrix ${{{\bf{Y}}^{\left( i \right)}_{{\rm{DD}}}}}$ of the $i$-th user, i.e., ${{\bf{y}}^{\left( i \right)}_{{\rm{DD}}}} \buildrel \Delta \over = {\rm{vec}}\left( {{{\bf{Y}}^{\left( i \right)}_{{\rm{DD}}}}} \right)$, and ${\tilde g_{l,l_p^{\left( i \right)},k,k_p^{\left( i \right)},p}^{\left( {i,j} \right)}}$ characterizes the symbol-wise effective channel coefficient, including the angular domain interference from the $j$-th user/beam to the $i$-th user/beam, the fading coefficient from the $p$-th path of the $i$-th user, and the phase rotation due to the twisted convolution, and is given by{\footnote{The additional phase term in the second line of~\eqref{symbol_wise_coefficient} is the consequence of the quasi-periodicity of the Zak transform~\cite{janssen1988zak}.}}
\begin{align}
\tilde g_{l,l_p^{\left( i \right)},k,k_p^{\left( i \right)},p}^{\left( {i,j} \right)} \!=\! \left\{ \begin{array}{l}
{\sqrt {{N_{{\rm{BS}}}}} }g_p^{\left( i \right)}\left[ j \right]h_p^{\left( i \right)}\!\exp \left( {j2\pi \frac{{k_p^{\left( i \right)}\left( {l - l_p^{\left( i \right)}} \right)}}{{MN}}} \right)\quad\quad\quad\quad\quad\quad\quad\quad\quad \!\! \!, l - l_p^{\left( i \right)} \!\ge\! 0\\
{\sqrt {{N_{{\rm{BS}}}}} }g_p^{\left( i \right)}\left[ j \right]h_p^{\left( i \right)}\!\exp \left( {j2\pi \frac{{k_p^{\left( i \right)}\left( {l - l_p^{\left( i \right)}} \right)}}{{MN}}} \right)\!\exp \left( { - j2\pi \frac{{\left( {k - k_p^{\left( i \right)}} \right)}}{N}} \right), l - l_p^{\left( i \right)} \!<\! 0
\end{array} \right.
. \label{symbol_wise_coefficient}
\end{align}

To further characterize the interference pattern, let us assume that the channel strengths, i.e., absolute values of fading coefficients, associated to each user are sorted in descending order, i.e., $\left| {h_1^{\left( i \right)}} \right| \ge \left| {h_2^{\left( i \right)}} \right| \ge ... \ge \left| {h_P^{\left( i \right)}} \right|$, for $1 \le i \le K$, without loss of generality.
In this case, the BS forms multi-beams towards the directions of the first paths of all users.
We henceforth refer to the first path of each user as the BF path, while the other paths are called non-BF paths.  With these in mind, we can expand~\eqref{X_Y_i_o_relationship} to yield
\begin{align}
Y_{{\rm{DD}}}^{\left( i \right)}\left[ {l,k} \right] = &\underbrace {\tilde g_{l,l_1^{\left( i \right)},k,k_1^{\left( i \right)},1}^{\left( {i,i} \right)}X_{{\rm{DD}}}^{\left( i \right)}\left[ {{{\left[ {l - l_1^{\left( i \right)}} \right]}_M},{{\left[ {k - k_1^{\left( i \right)}} \right]}_N}} \right]}_{{\rm{Desired \; signal}}} + \notag\\
&\underbrace {\sum\limits_{p = 2}^P {\tilde g_{l,l_p^{\left( i \right)},k,k_p^{\left( i \right)},p}^{\left( {i,i} \right)}X_{{\rm{DD}}}^{\left( i \right)}\left[ {{{\left[ {l - l_p^{\left( i \right)}} \right]}_M},{{\left[ {k - k_p^{\left( i \right)}} \right]}_N}} \right]} }_{{\rm{MPSI}}} + \notag\\
&\underbrace {\sum\limits_{\scriptstyle j = 1\hfill\atop
\scriptstyle j \ne i\hfill}^K {\tilde g_{l,l_1^{\left( i \right)},k,k_1^{\left( i \right)},1}^{\left( {i,j} \right)}X_{{\rm{DD}}}^{\left( j \right)}\left[ {{{\left[ {l - l_1^{\left( i \right)}} \right]}_M},{{\left[ {k - k_1^{\left( i \right)}} \right]}_N}} \right]} }_{{\rm{IBI}}} + \notag\\
&\underbrace {\sum\limits_{p = 2}^P {\sum\limits_{\scriptstyle j = 1\hfill\atop
\scriptstyle j \ne i\hfill}^K {\tilde g_{l,l_p^{\left( i \right)},k,k_p^{\left( i \right)},p}^{\left( {i,j} \right)}X_{{\rm{DD}}}^{\left( j \right)}\left[ {{{\left[ {l - l_p^{\left( i \right)}} \right]}_M},{{\left[ {k - k_p^{\left( i \right)}} \right]}_N}} \right]} } }_{{\rm{CTI}}} + {{{w}}^{\left( i \right)}\left[ {l,k} \right]} .\label{symbol_wise_coefficient_expanded}
\end{align}
From~\eqref{symbol_wise_coefficient_expanded}, we notice that the value of $Y_{{\rm{DD}}}^{\left( i \right)}\left[ {l,k} \right]$ is composed of several terms with different physical meanings. We can characterize those signals based on their physical meanings as follows:
\begin{itemize}
\item Desired signal: The first term in~\eqref{symbol_wise_coefficient_expanded} is the desired signal. The desired signal contains the information of the desired user and it is transmitted from the BF path.
\item MPSI: The second term in~\eqref{symbol_wise_coefficient_expanded} is the MPSI. The MPSI contains the interference from the desired user caused by the multi-path transmissions from the non-BF paths of the desired user.
\item IBI: The third term in~\eqref{symbol_wise_coefficient_expanded} is the IBI. The IBI contains the interference from other users caused by the superposition among different beams, as each user has a distinctive beam.
\item CTI: The fourth term in~\eqref{symbol_wise_coefficient_expanded} is the CTI. The CTI contains the interference from other users caused by the unintended alignment between the other users' BF directions and the desired user's non-BF paths.
\end{itemize}
A brief diagram characterizing the interference pattern is given in Fig.~\ref{THP_interference_model}, where both the IBI and CTI are clearly indicated.
As implied by the interference descriptions above, we notice that the interference terms have different characteristics. However, it should be noted that not all those interference terms make a significant contribution to the received symbol $Y_{{\rm{DD}}}^{\left( i \right)}\left[ {l,k} \right]$. In particular, user scheduling is usually performed at the BS before transmitting the downlink signals. One of the objectives of performing user scheduling is to avoid severe interference among different beams, which is enabled by grouping users with diverse spatial characteristics, e.g., AoDs~\cite{nam2014joint}. Furthermore, thanks to the nature of BF, the impact of MPSI is generally small. This is because the BS only forms narrow beams towards the BF paths of each user, and consequently the residual power on the non-BF paths is low. However, it can be shown that the CTI could have a high impact if the BF path of one user overlaps with one of the non-BF paths from a different user. This is because the transmitted signal after BF usually has a large power towards the BF direction. Therefore, even though the non-BF path may not have a large channel gain, the overall received power is still non-negligible as the transmitted power towards this direction is large.

\begin{figure}
\centering
\includegraphics[width=0.8\textwidth]{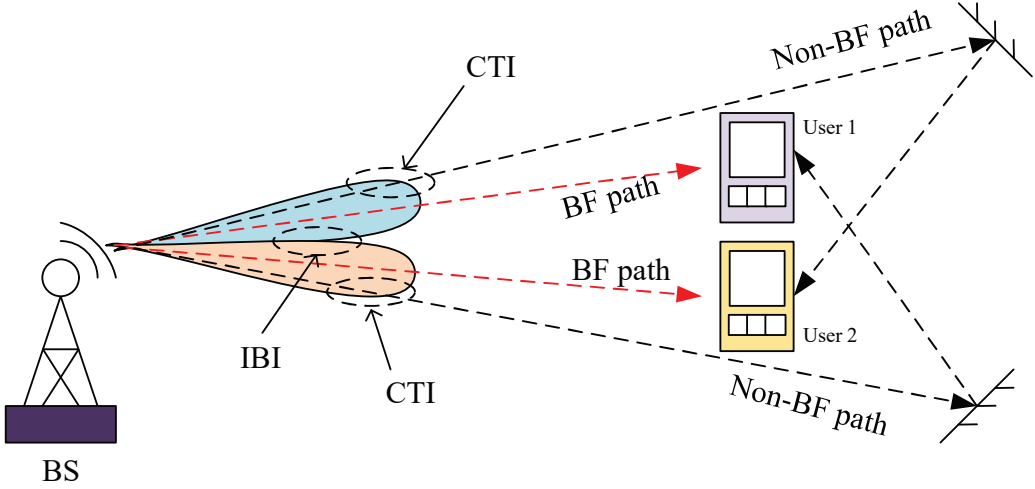}
\caption{The brief diagram of the interference pattern for downlink MU-MIMO-OTFS transmissions.}
\label{THP_interference_model}
\centering
\end{figure}

\subsection{Approximations with User Grouping}
As indicated by the discussions in the previous subsection, the interference terms have different characteristics. In the following subsection, we will develop a DD domain THP scheme by exploiting the nature of those interference terms with the aid of user grouping.
Let us consider the following assumption for user grouping:
\begin{itemize}
\item \textbf{Assumption 1}: We assume that the beams formulated for different users in the group are sufficiently separated (orthogonal) in the angular domain by having $N_{\rm BS} \gg K$. With this assumption, it is reasonable to ignore the IBI between different users. 
\end{itemize}
Furthermore, it should be noted that the AoDs of different paths associated to the same user are usually separated, especially for a sufficiently large number of transmit antennas. On top of that, the non-BF paths usually have much lower channel gain compared to the BF paths in practical settings thanks to the BF. Those two observations give rise to the following assumption:
\begin{itemize}
\item \textbf{Assumption 2}: We assume that the non-BF paths associated to the same user are relatively separated in the angular domain, where the channel gains are much lower compared to that of the BF path. With this assumption, it is reasonable to ignore the MPSI of each user.
\end{itemize}
We henceforth refer to the transmission where both assumptions 1 and 2 hold as the favorable propagation conditions, which is realizable with $N_{\rm BS} \gg K$.
Under the favorable propagation conditions,~\eqref{symbol_wise_coefficient_expanded} becomes
\begin{align}
Y_{{\rm{DD}}}^{\left( i \right)}\left[ {l,k} \right] \approx &\tilde g_{l,l_1^{\left( i \right)},k,k_1^{\left( i \right)},1}^{\left( {i,i} \right)}X_{{\rm{DD}}}^{\left( i \right)}\left[ {{{\left[ {l - l_1^{\left( i \right)}} \right]}_M},{{\left[ {k - k_1^{\left( i \right)}} \right]}_N}} \right] +\notag\\
&\sum\limits_{p = 1}^L {\tilde g_{l,l_{{P_i}\left[ p \right]}^{\left( i \right)},k,k_{{P_i}\left[ p \right]}^{\left( i \right)},{P_i}\left[ p \right]}^{\left( {i,{B_i}\left[ p \right]} \right)}} X_{{\rm{DD}}}^{\left( {{B_i}\left[ p \right]} \right)}\left[ {{{\left[ {l - l_{{P_i}\left[ p \right]}^{\left( i \right)}} \right]}_M},{{\left[ {k - k_{{P_i}\left[ p \right]}^{\left( i \right)}} \right]}_N}} \right] + {w^{\left( i \right)}}\left[ {l,k} \right]
, \label{X_Y_i_o_relationship_appro}
\end{align}
where the MPSI, IBI are ignored and only $L$ CTI terms are considered with $1 \le L \le \left( {P - 1} \right)\left( {K - 1} \right)$.
Here, the term $L$ is the number of CTI terms with significant power that will be considered in the precoding. The introduction of $L$ aims to strike a balance between the error performance and the computational complexity of the precoder. In~\eqref{X_Y_i_o_relationship_appro}, we define ${\bf B}_i$ of length $L$ as the \emph{CTI beam vector} for the $i$-th user and ${\bf P}_i$ of length $L$ as the \emph{CTI path vector} for the $i$-th user, respectively. The CTI beam vector contains the beam indices that correspond to the $L$ CTI terms with the most significant power for the $i$-th user, while the CTI path vector contains the indices of paths for the $i$-th user that spatially overlap with the beams with indices given in the CTI beam vector. In other words, with a descending power order of the CTI terms, the $p$-th CTI term, for $1 \le p \le L$, is caused by ${{B_i}\left[ p \right]}$-th beam overlaping with the ${{P_i}\left[ p \right]}$-th path of the $i$-th user. In particular, by examining~\eqref{symbol_wise_coefficient}, the elements of ${\bf B}_i$ and ${\bf P}_i$ can be determined based on the absolute values of ${h_i}\left[ {p} \right]g_{p}^{\left( i \right)}\left[ j \right]$, for $2 \le p \le P$ and $1 \le j  \le K$, $j \ne i$.


The approximated input-output relation in~\eqref{X_Y_i_o_relationship_appro} has an important property. For each DD domain received symbol, all the related DD domain transmitted symbols that contribute to the interference of this received symbol are from different DD grids of other users, as indicated in Fig.~\ref{OTFS_vs_OFDM_1}. This is quite different from the OFDM counterpart, where all the related TF domain transmitted symbols that contribute to a specific received TF domain symbol are from the same TF grid of different users, as indicated in Fig.~\ref{OTFS_vs_OFDM_2}. The rationale behind this observation is that the TF domain channel operation can be characterized by an element-wise product~\cite{tse2005fundamentals}, while the DD domain channel operation is characterized by the twisted convolution~\cite{lampel2021orthogonal}.
In fact, this property is the key enabler for a reduced-complexity THP for downlink MU-MIMO transmissions, which will be introduced in detail in the coming subsection.
\begin{figure}[htbp]
\setcounter{subfigure}{0}
\centering
\subfigure[MU-MIMO-OTFS transmission.]{
\begin{minipage}[t]{0.5\textwidth}
\centering
\includegraphics[scale=0.4]{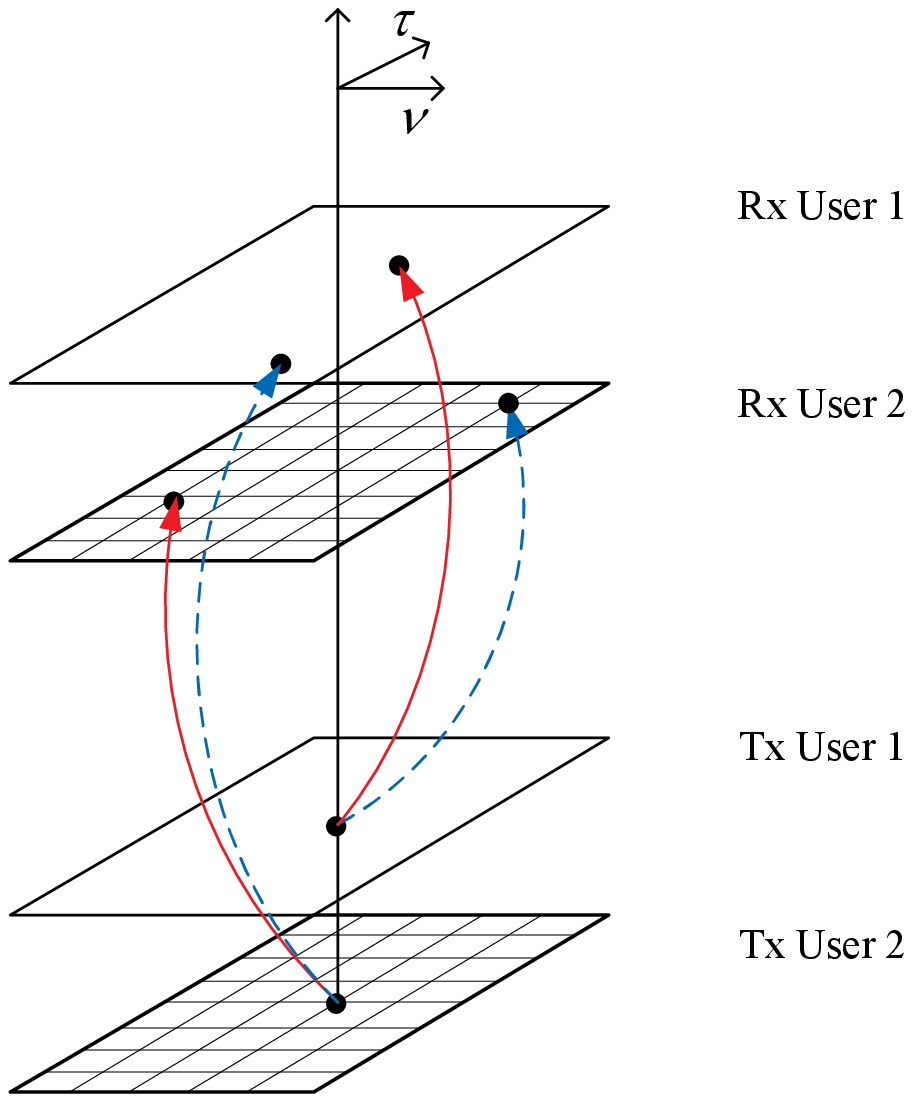}
\label{OTFS_vs_OFDM_1}
\end{minipage}%
}%
\centering
\subfigure[MU-MIMO-OFDM transmission.]{
\begin{minipage}[t]{0.5\textwidth}
\centering
\includegraphics[scale=0.4]{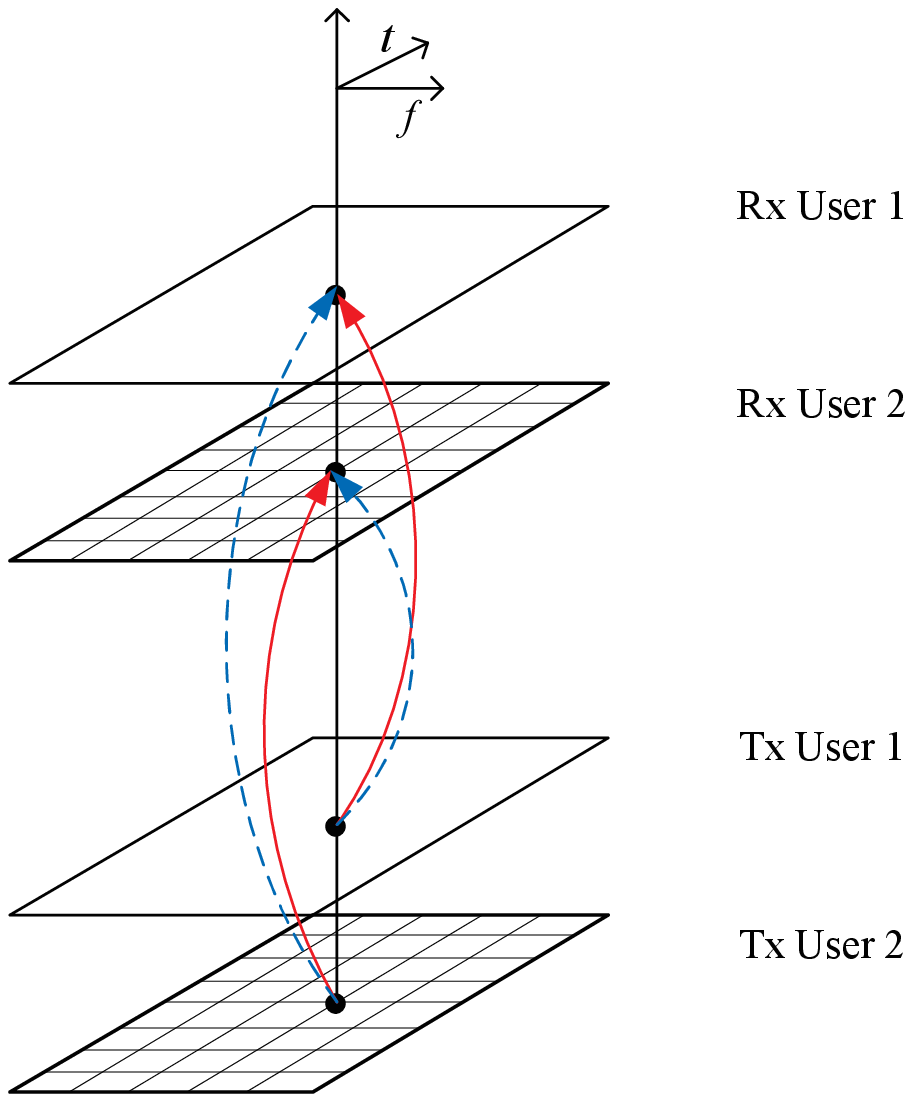}
\label{OTFS_vs_OFDM_2}
\end{minipage}%
}
\caption{A diagram characterizing the difference of interference patterns between MU-MIMO-OTFS and MU-MIMO-OFDM, where two users are considered. In particular, the red arrow denotes the BF path, while the blue dashed line implies the CTI.}
\label{OTFS_vs_OFDM}
\end{figure}

\subsection{DD Domain THP}
The core idea of THP is to pre-cancel the interference before transmission, where a modulo operation is applied to control the transmitted signal power~\cite{Harashima1972matched,Wesel1998Achievable}.
Before introducing the considered DD domain THP, let us consider the following example as shown in Fig.~\ref{THP_example_K2_P2}, where $M=N=3$, $P=2$, and $K=2$, respectively. There are in total $9$ DD grids for each user and we use the capital letters $A$ to $I$ to refer to the DD domain transmitted symbols associated to each DD grid in the ``Transmitter'' part, where the subscripts for the capital letters denote the corresponding user indices.
Furthermore, we use the solid and dashed arrows indicating the resolvable paths with different DD shifts, where we assume
that $l_1^{\left( 1 \right)} = 0,k_1^{\left( 1 \right)} = 0$, and $l_2^{\left( 1 \right)} = 0,k_2^{\left( 1 \right)} = -1$ for user 1, while $l_1^{\left( 2 \right)} = 1,k_1^{\left( 2 \right)} = 0$, and $l_2^{\left( 2 \right)} = 0,k_2^{\left( 2 \right)} = 1$ for user 2 as indicated by the bar chart attached to each path. Here, we assume that the positive delay and Doppler indices shift the symbol up and to the left, while the negative delay and Doppler indices shift the symbol down and to the right. The interference pattern corresponding to~\eqref{X_Y_i_o_relationship_appro} is shown in the ``Receiver'' part of Fig.~\ref{THP_example_K2_P2}, where the symbols on the left hand side in each DD grid is the desired signal (same color as the corresponding BF path), while the symbols on the right hand side are the interference (same color as the corresponding non-BF path).

\begin{figure}
\centering
\includegraphics[width=0.6\textwidth]{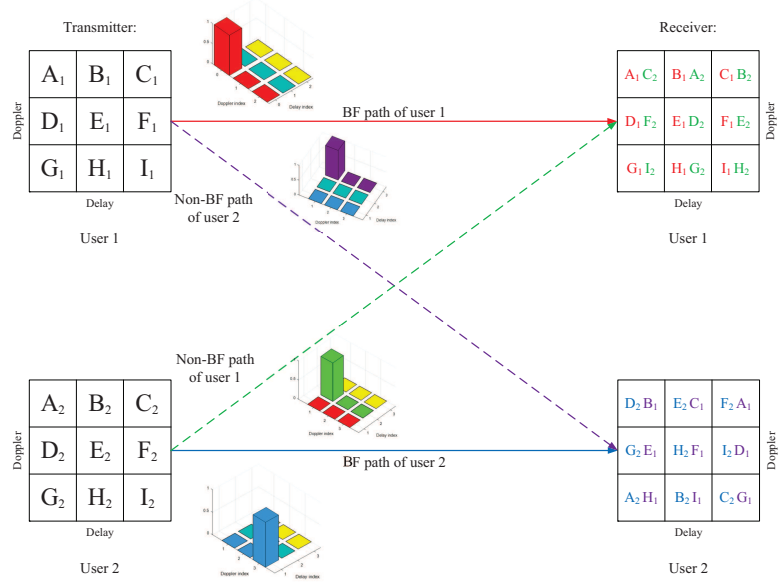}
\caption{An example of the interference pattern for MU-MIMO-OTFS, where $M=N=3$, $P=2$, and $K=2$, respectively.}
\label{THP_example_K2_P2}
\centering
\vspace{-5mm}
\end{figure}

It is interesting to note from Fig.~\ref{THP_example_K2_P2} that there is a possibility that we can directly pre-cancel all the interference in the DD domain by exploiting the different delay and Doppler responses associated to different paths.
For example, the received value of the first DD grid for user 1 only consists of the desired signal $A_1$ and the interference from $C_2$. Therefore, the
interference for $A_1$ can be perfectly canceled if we know the exact value of $C_2$. Similarly, the interference for $C_2$ can be canceled if we know the exact value of $G_1$. So on and so forth, it can be shown that there are DD domain cycles that contain several DD domain symbols for the interference cancellation, e.g., $A_1 \to C_2\to G_1\to I_2 \to D_1 \to F_2 \to A_1$. However, it should be noted that the pre-cancellation could change the value of the corresponding DD domain transmitted symbols. Consequently, due to the DD domain cycles, the pre-cancellation of interference cannot be directly applied. For instance, in the considered example, to pre-cancel the interference for $A_1$, it is required to know the value of $A_1$ after interference cancellation as suggested by the cycle, which is a non-causal operation and cannot be implemented in practice.

To solve this problem, we propose to assign known symbols to specific DD grids in order to break the DD domain cycles. For example, if we assign a zero to the symbol $A_1$, then the pre-cancellation for $F_2$ can be conducted. Following the DD domain cycle, the interference can be pre-cancelled step by step, such as $A_1 \to F_2\to D_1\to I_2 \to G_1 \to C_2$. The corresponding pre-cancelation is illustrated in Fig.~\ref{THP_example_K2_P2_cancellation}, where there are in total $3$ DD domain cycles. We use superscripts with different numbers and colors to represent the schedule of interference cancellation for each DD domain cycle, where we set $A_1$, $D_2$, and $C_1$ as zeros. The zeros in superscript represent the initialization for the pre-cancelation of the corresponding DD domain cycle, while the ones in superscript mark the start of the pre-cancelation.
It is not hard to see that the considered pre-cancellation can indeed cancel all the interference without any matrix decomposition or inversion via intentionally assigning known symbols.
\begin{figure}
\centering
\includegraphics[width=0.6\textwidth]{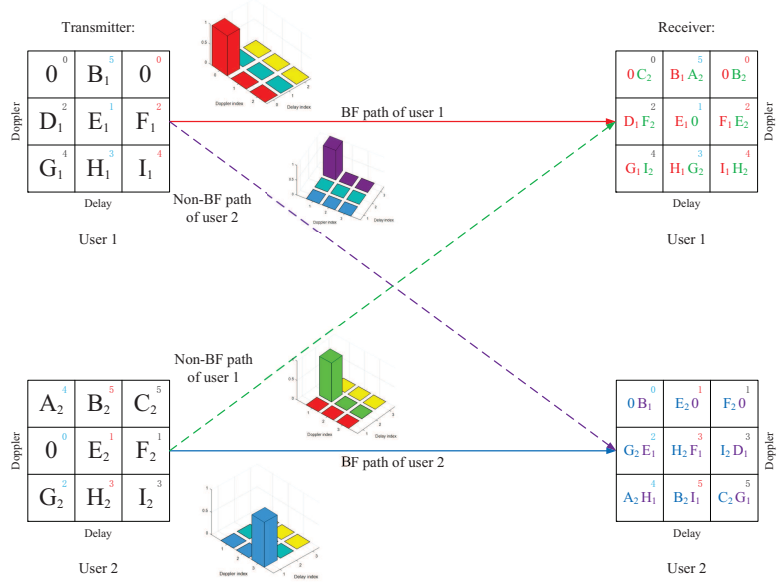}
\caption{The application of DD domain THP for the example given in Fig.~\ref{THP_example_K2_P2}.}
\label{THP_example_K2_P2_cancellation}
\centering
\vspace{-5mm}
\end{figure}

Based on the above example, we are ready to present the implementation of DD domain THP. Note that the proposed THP follows a symbol-by-symbol pre-cancelation, and for each DD domain symbol, it is required to know where the interference comes from and which symbol should be pre-canceled next.
Let us denote by ${{\bf{\hat B}}}$ of length $K$ the \emph{interfered beam vector} for all the users and ${{\bf{\hat P}}}$ of length $K$ the \emph{interfered path vector} for all the users. In particular, the $i$-th element of ${{\bf{\hat B}}}$ is the index of the user, to whom the $i$-th beam (the transmitted signal of the $i$-th user) causes the most significant CTI, and the $i$-th element of ${{\bf{\hat P}}}$ is the corresponding path index, from which the ${{{\hat B}}}[i]$-th user receives the CTI due to the $i$-th beam.
Those terms indicate the precoding schedule for the considered THP scheme, as the most significant CTI from the $i$-th beam is likely to be included in the CTI beam vector of the $\hat B\left[ i \right]$-th user. In this case, the symbols in the $i$-th beam after pre-cancellation are likely to be used for the pre-cancellation for the $\hat B\left[ i \right]$-th user, thereby reducing the overhead.
In particular, by observing~\eqref{symbol_wise_coefficient},
we have $\hat B\left[ i \right] \buildrel \Delta \over = \arg \mathop {\max }\limits_i \left| {{h_j}\left[ p \right]g_p^{\left( j \right)}\left[ i \right]} \right|$, for $2 \le p \le P$ and $1 \le j \le K$, $j\le i$, and $\hat P\left[ i \right] \buildrel \Delta \over = \arg \mathop {\max }\limits_p \left| {{h_{\hat B\left[ i \right]}}\left[ p \right]g_p^{\left( {\hat B\left[ i \right]} \right)}\left[ i \right]} \right|$, for $2 \le p \le P$.
Corresponding to the above discussions, the details of DD domain THP are summarized in Algorithm~1, where $\bmod \left[ \cdot \right]$ denotes the modulo operation in the conventional THP. Some discussions on the modulo threshold will be presented in the coming section.

As implied by Algorithm~1, the $L$ most significant CTI will be pre-cancelled via THP for each DD domain symbol. Therefore, according to~\eqref{X_Y_i_o_relationship_appro} and the principle of THP, the receiver side applies a single-tap equalization together with a modulo operation to recover the DD domain transmitted symbols~\cite{Wesel1998Achievable}. In particular, we have
\begin{align}
\hat Y_{{\rm{DD}}}^{\left( i \right)}\left[ {l,k} \right] = \bmod \left[ {\frac{1}{{\tilde g_{l,l_1^{\left( i \right)},k,k_1^{\left( i \right)},1}^{\left( {i,i} \right)}}}Y_{{\rm{DD}}}^{\left( i \right)}\left[ {l,k} \right]} \right]. \label{Receiver_single_tap_mod}
\end{align}
Based on $\hat Y_{{\rm{DD}}}^{\left( i \right)}\left[ {l,k} \right]$, a straightforward demodulation could be applied to recover the transmitted information for each user.

\begin{algorithm}[htb]
\small
\caption{DD Domain THP for Downlink MU-MIMO-OTFS Transmissions}
\hspace*{0.02in} {\bf Input:}
$\tilde g_{l,l_p^{\left( i \right)},k,k_p^{\left( i \right)},p}^{\left( {i,j} \right)},{\bf{S}}_{{\rm{DD}}}^{\left( i \right)},l_p^{\left( i \right)},k_p^{\left( i \right)}$, ${\bf P}_i$,
${\bf B}_i$, ${\bf \hat B}$, and ${\bf \hat P}$,\\
\hspace*{0.5in}for $0 \le l \le M-1$, $0 \le k \le N-1$, $1 \le p \le P$, $1 \le i,j \le K$.\\
\hspace*{0.02in} {\bf Initialization:}
Set ${\rm{Indicator\_mtx}}[l,k,i] = 0$, for $0 \le l \le M-1$, $0 \le k \le N-1$, $1 \le i \le K$.\\
\hspace*{0.93in}Set ${\rm{Overhead\_mtx}}[l,k,i] = 0$, for $0 \le l \le M-1$, $0 \le k \le N-1$, $1 \le i \le K$.\\
\hspace*{0.02in} {\bf Steps:}
\begin{algorithmic}[1]
\State \textbf{for} $l'$ from $0$ to $M-1$ \textbf{do}
\State $\quad$\textbf{for} $k'$ from $0$ to $N-1$ \textbf{do}
\State $\quad$$\quad$\textbf{for} $i'$ from $1$ to $K$ \textbf{do}
\State $\quad$$\quad$$\quad$ Set $l= l'$, $k= k'$, and $i= i'$.
\State $\quad$$\quad$$\quad$ \textbf{while} ${\rm{Indicator\_mtx}}[{l},{k},{i}] = 0$ \textbf{do}
\State $\quad$$\quad$$\quad$$\quad$ $X_{{\rm{DD}}}^{\left( {i} \right)}\left[ {l,k} \right] = S_{{\rm{DD}}}^{\left( {i} \right)}\left[ {l,k} \right]$.
\State $\quad$$\quad$$\quad$$\quad$ \textbf{for} $p$ from $1$ to $L$ \textbf{do}
\State $\quad$$\quad$$\quad$$\quad$$\quad$ Set ${\rm{delay}}\_{\rm{idx}} = {\left[ {{{\left[ {{l} - l_1^{\left( {i} \right)}} \right]}_M} + l_{P_i[p]}^{\left( i \right)}} \right]_M}$ and ${\rm{Doppler}}\_{\rm{idx}} = {\left[ {{{\left[ {{k} - k_1^{\left( {i} \right)}} \right]}_N} + k_{P_i[p]}^{\left( i \right)}} \right]_N}$.
\State $\quad$$\quad$$\quad$$\quad$$\quad$ \textbf{if} ${\rm{Indicator\_mtx}}[{\rm{delay}}\_{\rm{idx}},{\rm{Doppler}}\_{\rm{idx}},B_i[p]] = 0$ \textbf{do}
\State $\quad$$\quad$$\quad$$\quad$$\quad$$\quad$  Set ${{ X}}_{{\rm{DD}}}^{\left( B_i[p] \right)}[{\rm{delay}}\_{\rm{idx}},{\rm{Doppler}}\_{\rm{idx}}]=0 $.
\State $\quad$$\quad$$\quad$$\quad$$\quad$$\quad$  Set ${\rm{Indicator\_mtx}}[{\rm{delay}}\_{\rm{idx}},{\rm{Doppler}}\_{\rm{idx}},B_i[p]] = 1$.
\State $\quad$$\quad$$\quad$$\quad$$\quad$$\quad$  Set ${\rm{Overhead\_mtx}}[{\rm{delay}}\_{\rm{idx}},{\rm{Doppler}}\_{\rm{idx}},B_i[p]] = 1$.
\State $\quad$$\quad$$\quad$$\quad$$\quad$  \textbf{end if}
\State $\quad$$\quad$$\quad$$\quad$$\quad$ $X_{{\rm{DD}}}^{\left( {i} \right)}\left[ {l,k} \right] = X_{{\rm{DD}}}^{\left( {i} \right)}\left[ {l,k} \right] - \frac{{\tilde g_{l,l_{P_i[p]}^{\left( {i} \right)},k,k_{P_i[p]}^{\left( {i} \right)},P_i[p]}^{\left( {i,B_i[p] } \right)}}}{{\tilde g_{ l,l_1^{\left( {i} \right)},k,k_1^{\left( {i} \right)},1}^{\left( {i,i} \right)}}}{ X}_{{\rm{DD}}}^{\left( B_i[p]  \right)}\left[ {{\rm{delay}}\_{\rm{idx}},{\rm{Doppler}}\_{\rm{idx}}} \right]$.
\State $\quad$$\quad$$\quad$$\quad$ \textbf{end for}
\State $\quad$$\quad$$\quad$$\quad$ $ X_{{\rm{DD}}}^{\left( i \right)}\left[ {l, k} \right] = \bmod \left[ { X_{{\rm{DD}}}^{\left( i \right)}\left[ {l, k} \right]} \right]$.
\State $\quad$$\quad$$\quad$$\quad$ Set ${\rm{Indicator\_mtx}}[l,k,i] = 1$.
\State $\quad$$\quad$$\quad$$\quad$ Set $l = {\left[ {{{\left[ {l - l_{\hat P\left[ i \right]}^{\left( {\hat B\left[ i \right]} \right)}} \right]}_M} + l_1^{\left( {\hat B\left[ i \right]} \right)}} \right]_M}$, $k = {\left[ {{{\left[ {k - k_{\hat P\left[ i \right]}^{\left( {\hat B\left[ i \right]} \right)}} \right]}_N} + k_1^{\left( {\hat B\left[ i \right]} \right)}} \right]_N}$, and $i = \hat B\left[ i \right]$.
\State $\quad$$\quad$$\quad$ \textbf{end while}
\State $\quad$$\quad$\textbf{end for}
\State $\quad$\textbf{end for}
\State \textbf{end for}
\State \textbf{Return} ${{\hat X}}_{{\rm{DD}}}^{\left( i \right)}$, for $1 \le i \le K$.
\end{algorithmic}
\label{DD_Domain_THP}
\end{algorithm}

\subsection{Complexity and Signaling Overhead}
We will discuss the computational complexity and the required signaling overhead for the considered THP in this subsection. As indicated by Algorithm~1, there are at most $L$ times of pre-cancellation for each DD domain transmitted symbol. Thus, the overall computational complexity is linear to the number of transmitted symbols with a linearity coefficient $L$, i.e., ${\cal O}\left( LKMN \right)$. It should be noted that such a linear complexity is lower than most of the existing precoding schemes for MU-MIMO-OTFS, including the ones in~\cite{Ramachandran2018mimo,Pandey2021low}, because the proposed THP does not rely on the complex channel decomposition or inversion.

On the other hand, it can be observed that the signaling overhead for the proposed THP depends on the value of $L$, and the channel conditions, such as the number of paths, number of users, and delay and Doppler responses. Furthermore, the pre-cancellation order is also of great importance for the signaling overhead. Note that Algorithm~1 is a \emph{performance-centric} implementation of DD domain THP, where the algorithm aims to pre-cancel all the interference terms without considering the required overhead. Consequently, the total number of assigned known symbols increases if the corresponding interference symbols have not yet been pre-cancelled, e.g., line 9 to 13 in Algorithm~1. In contrast, there could also be an \emph{overhead-centric} implementation, where the pre-cancellation is performed with the priority to the symbols, to whom the corresponding interference symbols have already been pre-cancelled, e.g., line 14 in Algorithm~1, in order to minimized the required overhead.
However, the reduced overhead implementation is currently still an open problem and
we are unable to discuss this issue in detail due to the space limitation. But it should be pointed out that the searching algorithms for tree- and trellis-based graphical models may shed light on this issue~\cite{Bocharova2004beast,li2017reduced}.

\section{Achievable Rate Analysis}
\begin{figure}[tbp]
\setcounter{subfigure}{0}
\centering
\subfigure[Equivalent diagram of the system model in Fig.~\ref{System_Model_MU_MIMO}.]{
\begin{minipage}[t]{0.5\textwidth}
\centering
\includegraphics[scale=0.5]{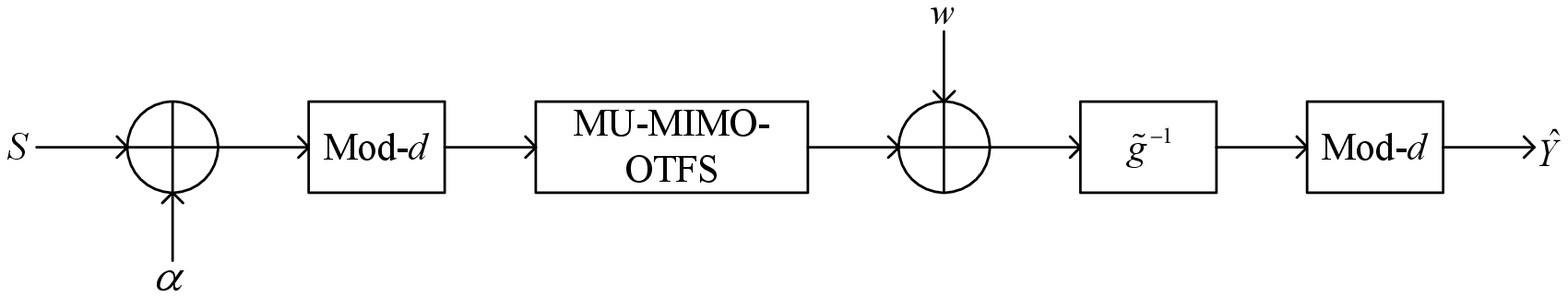}
\label{equivalent_diagram1}
\end{minipage}%
}%
\centering
\subfigure[Simplified diagram of Fig.~\ref{equivalent_diagram1}.]{
\begin{minipage}[t]{0.5\textwidth}
\centering
\includegraphics[scale=0.5]{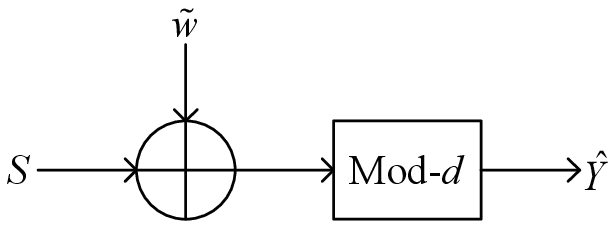}
\label{equivalent_diagram2}
\end{minipage}%
}
\caption{Equivalent and simplified system models corresponding to Fig.~\ref{System_Model_MU_MIMO}.}
\label{equivalent_diagrams}
\vspace{-5mm}
\end{figure}

We discuss the achievable rates of the proposed THP scheme in this section. Without loss of generality, we consider the quadrature amplitude modulation (QAM) constellation set{\footnote{Although we only focus on QAM constellation here, the related discussions can be straightforwardly extended to the case of general constellations, e.g., pulse amplitude modulation (PAM).}} $\mathbb A$.
In particular, we focus on the average achievable rate for each DD domain symbol under favorable propagation conditions by assuming that $N_{\rm BS} \gg K$.
For ease of derivation, we
provide an equivalent diagram of the proposed THP-based MU-MIMO-OTFS characterizing the corresponding processing between $S_{{\rm{DD}}}^{\left( i \right)}\left[ {l,k} \right]$  and ${\hat Y}_{{\rm{DD}}}^{\left( i \right)}\left[ {{{\left[ {l + l_1^{\left( i \right)}} \right]}_M},{{\left[ {k + k_1^{\left( i \right)}} \right]}_N}} \right]$ in Fig.~\ref{equivalent_diagram1}, where we neglect the symbol indices for notational brevity. Specifically, we use the term $\alpha$ in Fig.~\ref{equivalent_diagram1} to describe the pre-cancellation of THP. As indicated by this diagram, an arbitrary DD domain symbol $S$ after pre-cancellation with term $\alpha$ and modulo operation with threshold $d$ is transmitted over the MU-MIMO-OTFS channel. The received channel observation contains the corruption from the AWGN sample $w$, which is used for symbol detection after an single tap equalization with ${\tilde g}^{-1}$, e.g., ${\left( {\tilde g_{l,l_1^{\left( i \right)},k,k_1^{\left( i \right)},1}^{\left( {i,i} \right)}} \right)^{ - 1}}$, and applying the  modulo operation with threshold $d$. Those descriptions are consistent with our system model in Section II. In particular, the above processing can be described by the following equation
\begin{align}
\hat Y = \bmod \left[ {\frac{1}{{\tilde g}}\left( {\tilde g\left( {\bmod \left[ {S + \alpha } \right]} \right) + \eta  + w} \right)} \right]= \bmod \left[ {\bmod \left[ {S + \alpha } \right] + \frac{1}{{\tilde g}}\left( {\eta  + w} \right)} \right], \label{equivalent_model_S_Y}
\end{align}
where $\eta$ denotes the interference term due to the MU-MIMO-OTFS transmission as suggested in~\eqref{symbol_wise_coefficient_expanded}.
Note that $\bmod \left[ {\bmod \left[ a \right]+b} \right] = \bmod \left[ {a + b} \right]$. Thus,~\eqref{equivalent_model_S_Y} can be further simplified to
\begin{align}
\hat Y =\bmod \left[ {S + \alpha  + \frac{1}{{\tilde g}}\eta  + \frac{1}{{\tilde g}}w} \right]. \label{equivalent_model_S_Y_der1}
\end{align}
Furthermore, as implied by Line 14 of Algorithm~1, the interference term ${\eta  \mathord{\left/
 {\vphantom {\eta  {\tilde g}}} \right.
 \kern-\nulldelimiterspace} {\tilde g}}$ will be cancelled by pre-cancellation, e.g., term $\alpha$, with a sufficiently large number of $L$, in the case of user grouping and BF. Therefore, we can further approximate~\eqref{equivalent_model_S_Y_der1} by
\begin{align}
\hat Y  \approx  \bmod \left[ {S + \frac{1}{{\tilde g}}w} \right]. \label{equivalent_model_appro}
\end{align}
The corresponding diagram to~\eqref{equivalent_model_appro} is presented in Fig.~\ref{equivalent_diagram2}, where $\tilde w = \frac{1}{{\tilde g}}w$ denotes the equivalent AWGN sample with one-sided PSD ${{{N_0}} \mathord{\left/
 {\vphantom {{{N_0}} {{{\left| {\tilde g} \right|}^2}}}} \right.
 \kern-\nulldelimiterspace} {{{\left| {\tilde g} \right|}^2}}}$.

Now we focus on the achievable rate for the considered scheme based on~\eqref{equivalent_model_appro}. In particular, the mutual information between $S$ and $\hat Y$ is given by~\cite{cover2012elements,Wesel1998Achievable}
\begin{align}
I\left( {S;\hat Y} \right) \buildrel \Delta \over = h\left( {\hat Y} \right) - h\left( {\hat Y|S} \right) \approx h\left( {\bmod \left[ {S + \frac{1}{{\tilde g}}w} \right]} \right) - h\left( {\bmod \left[ {\frac{1}{{\tilde g}}w} \right]} \right). \label{MU_SY}
\end{align}
Notice that the modulo operation strictly limits the signal value from $\left[ { - \frac{d}{2},\frac{d}{2}} \right]$ for both the real and imaginary dimensions, and the maximum entropy probability distribution for a random variable with support constrained to an interval is the independent and identically distributed (i.i.d.) uniform distribution~\cite{cover2012elements}.
Thus, ~\eqref{MU_SY} can be approximately upper-bounded by
\begin{align}
I\left( {S;\hat Y} \right) \lesssim 2\log_2 \left( d \right) - h\left( {\bmod \left[ {\frac{1}{{\tilde g}}w} \right]} \right). \label{MU_SY_der1}
\end{align}
Note that the values of AWGN samples are generally small in the high SNR regime. Thus, in the high SNR regime (e.g., the real/imaginary part of the noise sample is within the range of $\left[ { - \frac{d}{2},\frac{d}{2}} \right]$),~\eqref{MU_SY_der1} can be shown to converge to~\cite{Wesel1998Achievable}
\begin{align}
I\left( {S;\hat Y} \right) \lesssim 2\log_2 \left( d \right) - h\left( {\frac{1}{{\tilde g}}w} \right)=2{\log _2}\left( d \right) - {\log _2}\left( {\pi e\frac{{{N_0}}}{{{{\left| {\tilde g} \right|}^2}}}} \right)={\log _2}\left( {\frac{{{d^2}{{\left| {\tilde g} \right|}^2}}}{{\pi e{N_0}}}} \right). \label{MU_SY_der2}
\end{align}
Based on~\eqref{MU_SY_der2}, we are ready to investigate the sum-rate performance for the considered THP scheme. Notice that there is no joint decoding among different users. Thus, with favorable propagation conditions, the sum-rate for the considered downlink MU-MIMO-OTFS can be formulated by
\begin{align}
{R_{{\rm{sum}}}} \buildrel \Delta \over = \sum\limits_{i = 1}^K {I\left( {S_{{\rm{DD}}}^{\left( i \right)}\left[ {l,k} \right];\hat Y_{{\rm{DD}}}^{\left( i \right)}\left[ {l,k} \right]} \right)}  = \sum\limits_{i = 1}^K {{{\log }_2}\left( {\frac{{{d^2}{{\left| {\tilde g_{l,l_1^{\left( i \right)},k,k_1^{\left( i \right)},1}^{\left( {i,i} \right)}} \right|}^2}}}{{\pi e{N_0}}}} \right)}. \label{MU_SY_sum_rate}
\end{align}
Furthermore, by substituting~\eqref{symbol_wise_coefficient} into~\eqref{MU_SY_sum_rate}, we have
\begin{align}
{R_{{\rm{sum}}}} = \sum\limits_{i = 1}^K {{{\log }_2}\left( {\frac{{{d^2}{N_{{\rm{BS}}}}{{\left| {h_1^{\left( i \right)}} \right|}^2}}}{{\pi e{N_0}}}} \right)} .
\label{MU_SY_sum_rate_der1}
\end{align}
As implied by~\eqref{MU_SY_sum_rate_der1}, the sum-rate is related to the choice of modulo threshold $d$. 
According to~\cite{Wesel1998Achievable}, the average power for transmitted symbol $X_{\rm DD}^{(i)}$ converges to $d^2/12$ and $d^2/6$ for PAM and QAM constellations, respectively. Thus, with QAM constellations, the total transmit power for a given time slot is $Kd^2/6$. Based on the total transmit power, we can define the SNR for the THP transmission by ${\rm{SNR}} \buildrel \Delta \over = \frac{{K{d^2}}}{{6{N_0}}}$.
Finally, we obtain the sum-rate at high SNRs by
\begin{align}
{R_{{\rm{sum}}}} = \sum\limits_{i = 1}^K {{{\log }_2}\left( {\frac{6}{{\pi e}}\frac{{{N_{{\rm{BS}}}}}}{K}{{\left| {h_1^{\left( i \right)}} \right|}^2}} {\rm SNR}\right)}  .
\label{MU_SY_sum_rate_finial}
\end{align}

Next, we discuss some important insights based on the previous analysis. In particular, we restrict ourselves to the high SNR regime, where the sum-rate is characterized by~\eqref{MU_SY_sum_rate_finial}. Let us first characterize the sum-rate gap of the proposed scheme to the optimal transmission scenario, where there is only one resolvable path between the BS and each user with sufficiently separated (orthogonal) angular features. The latter transmission scenario is optimal in the sense that it does not have neither MPSI, IBI, nor CTI, and therefore maximizes the throughput of the downlink transmission. The following lemma shows the sum-rate in the optimal transmission scenario.

\textbf{Lemma 2} (\emph{Optimal Sum-rate}):
In the optimal transmission scenario, where there is only one resolvable path between the BS and each user without IBI, the sum-rate is given by
\begin{align}
R_{{\rm{sum}}}^{{\rm{opt}}} = \sum\limits_{i = 1}^K {{{\log }_2}\left( {1 + \frac{{{N_{{\rm{BS}}}}}}{K}{{\left| {h_1^{\left( i \right)}} \right|}^2}{\rm{SNR}}} \right)}  . \label{Lemma2}
\end{align}

\emph{Proof}: By considering the uniform power allocation among different users,~\eqref{Lemma2} can be derived by following the capacity calculation for parallel Gaussian channels with independent noise~\cite{cover2012elements}. The detail derivations are omitted here due to the space limitation. $\hfill\blacksquare$

Based on Lemma 2, the following theorem characterizes the sum-rate gap between the proposed scheme and the optimal case in the high SNR regime.

\textbf{Theorem 1} (\emph{Shaping Loss}): For sufficiently large $L$ (perfect pre-cancellation of interference) and $N_{\rm BS} \gg K$, the proposed scheme only has a constant rate loss for each user compared to the optimal transmission scenario in the high SNR regime.

\emph{Proof}:
\begin{align}
R_{{\rm{sum}}}^{{\rm{opt}}} - {R_{{\rm{sum}}}} = \sum\limits_{i = 1}^K {{{\log }_2}\left( {\frac{{1 + \frac{{{N_{{\rm{BS}}}}}}{K}{{\left| {h_1^{\left( i \right)}} \right|}^2}{\rm{SNR}}}}{{\frac{6}{{\pi e}}\frac{{{N_{{\rm{BS}}}}}}{K}{{\left| {h_1^{\left( i \right)}} \right|}^2}{\rm{SNR}}}}} \right)}  \approx \sum\limits_{i = 1}^K {{{\log }_2}\left( {\frac{{\pi e}}{6}} \right)} ,
\end{align}
where the approximation holds in the high SNR regime. Note that $\frac{1}{2}{\log _2}\left( {\frac{{\pi e}}{6}} \right) \approx 0.255$, which is the well-known ``shaping loss'' for general PAM constellations in the THP literature. \hfill $\blacksquare$

As implied by Theorem 1, the proposed scheme can obtain a promising rate performance that only has a constant gap to the optimal transmission. As pointed out by~\cite{forney1992trellis}, this performance loss is the ``shaping loss'', which is caused by the peak limitation introduced by precoding. Next, we will discuss the growth rate of the sum-rate with respect to different parameters. The following theorem shows the scaling law of the proposed scheme.

\textbf{Theorem 2} (\emph{Scaling Law for Sum-rate}):
For sufficiently large $L$ (perfect pre-cancellation of interference) and $N_{\rm BS} \gg K$, the sum-rate of the proposed scheme scales linearly with the number of users $K$ under favorable propagation conditions at the asymptotically high SNRs.

\emph{Proof}: Based on~\eqref{MU_SY_sum_rate_finial}, we have
\begin{align}
\mathop {\lim }\limits_{{\rm{SNR}} \to \infty } \frac{{{R_{{\rm{sum}}}}}}{{{{\log }_2}\left( {{\rm{SNR}}} \right)}} = \mathop {\lim }\limits_{{\rm{SNR}} \to \infty }\frac{{\sum\limits_{i = 1}^K {{{\log }_2}\left( {\frac{6}{{\pi e}}\frac{{{N_{{\rm{BS}}}}}}{K}{{\left| {h_1^{\left( i \right)}} \right|}^2}} \right) + K{{\log }_2}\left( {{\rm{SNR}}} \right)} }}{{{{\log }_2}\left( {{\rm{SNR}}} \right)}} = K,
\end{align}
which indicates that the sum-rate growth is linear in $K$.$\hfill\blacksquare$

The conclusion in Theorem 2 is not unexpected. Note that the proposed scheme contains $N_{\rm BS}$ antennas and $K$ RF, where $N_{\rm BS} > K$. Thus, it can be shown that the degree-of-freedom (DoF) of the proposed scheme is limited by $K$ instead of $N_{\rm BS}$~\cite{rusek2012scaling}, which in fact determines the maximum sum-rate growth rate (the pre-log factor) as shown in Theorem~2. Next, we study the sum-rate performance with respect to the number of antennas at BS ${N_{{\rm{BS}}}}$.

\textbf{Theorem 3} (\emph{Sum-Rate vs. ${N_{{\rm{BS}}}}$}):
For sufficiently large $L$ (perfect pre-cancellation of interference) and $N_{\rm BS} \gg K$, the sum-rate of the proposed scheme for a given $K$ increases logarithmically with the number of antennas at BS under favorable propagation conditions.

\emph{Proof}: Based on~\eqref{MU_SY_sum_rate_finial}, we have
\begin{align}
\mathop {\lim }\limits_{{\rm{SNR}} \to \infty } \frac{{{R_{{\rm{sum}}}}}}{{{{\log }_2}\left( {{N_{{\rm{BS}}}}} \right)}} =\mathop {\lim }\limits_{{\rm{SNR}} \to \infty } \frac{{\sum\limits_{i = 1}^K {{{\log }_2}\left( {\frac{6}{{\pi e}}\frac{{{{\left| {h_1^{\left( i \right)}} \right|}^2}}}{K}{\rm{SNR}}} \right) + K{{\log }_2}\left( {{N_{{\rm{BS}}}}} \right)} }}{{{{\log }_2}\left( {{N_{{\rm{BS}}}}} \right)}} = K,
\end{align}
which indicates that the sum-rate growth increases logarithmically with ${{N_{{\rm{BS}}}}}$.$\hfill\blacksquare$

The conclusion in Theorem 3 aligns with Theorem 2. As the DoF is determined by the number of users $K$, a larger number of $N_{\rm BS}$ can only provide the SNR gain, which is consistent with the general conclusions for MU-MIMO~\cite{rusek2012scaling}. The correctness of the above theorems will be verified in the coming section.

\section{Numerical Results}
In this section, we will use numerical results to verify the effectiveness of the proposed schemes. We consider MU-MIMO-OTFS transmissions with $M=32$ and $N=16$, where we set the maximum delay and Doppler indices to $l_{\rm max}=5$ and $k_{\rm max}=7$, respectively. The delay and Doppler indices are assumed to be integer values unless otherwise specified. The fading coefficients are generated based on the exponential power delay profile with a path loss exponent of 2.76. The signal constellation is the quadrature phase shift keying (QPSK) constellation.
Furthermore, we present the results under both favorable propagation and practical channel conditions. For the favorable propagation case, the received signals are generated based on~\eqref{symbol_wise_coefficient_expanded}, where both the MPSI and IBI are ignored. For the practical case, the received signals are generated based on~\eqref{X_Y_i_o_relationship}, and a user grouping strategy is applied such that the maximum spatial correlation between different users is no larger than $0.1$, i.e., ${g_p^{\left( i \right)}\left[ j \right]} \le 0.1$, for $i \ne j$. Meanwhile, we assume that the different resolvable paths have AoDs that are at least 5 degrees away from each other.

\subsection{Numerical Results under Favorable Propagation Conditions}
\begin{figure*}
\centering
\subfigure[Sum-rate performance for $K=2$ and different $N_{\rm BS}$.]{
\begin{minipage}[t]{0.5\textwidth}
\centering
\includegraphics[scale=0.5]{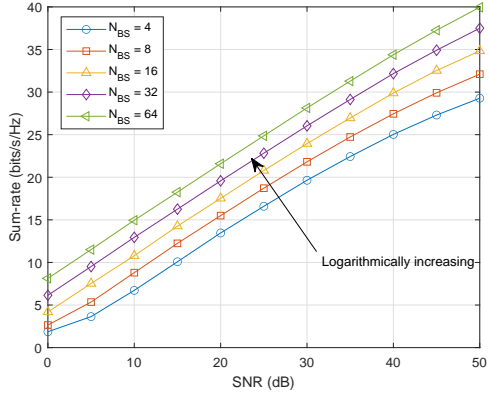}
\label{Sum_rate_vs_N_BS}
\end{minipage}%
}%
\subfigure[Sum-rate performance for different $K$ and $N_{\rm BS}$.]{
\begin{minipage}[t]{0.5\textwidth}
\centering
\includegraphics[scale=0.5]{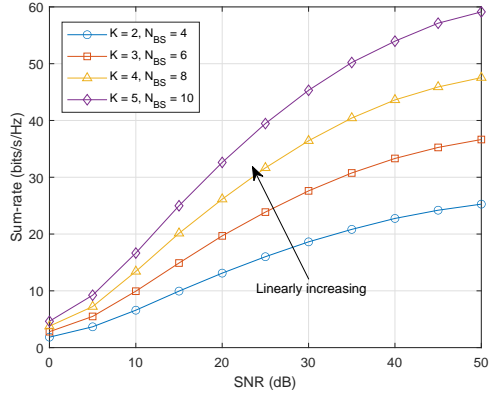}
\label{Sum_rate_vs_K}
\end{minipage}%
}%
\quad
\subfigure[Sum-rate performance for different values of $L$.]{
\begin{minipage}[t]{0.5\textwidth}
\centering
\includegraphics[scale=0.5]{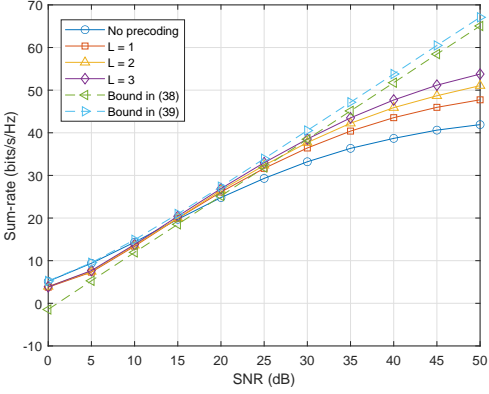}
\label{Sum_rate_vs_L}
\end{minipage}%
}%
\subfigure[BER performance for different $K$, $N_{\rm BS}$, and $L$.]{
\begin{minipage}[t]{0.5\textwidth}
\centering
\includegraphics[scale=0.5]{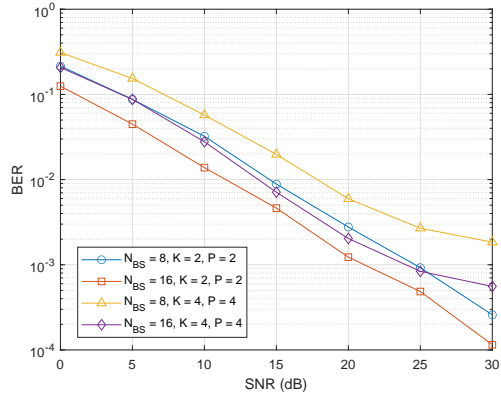}
\label{BER}
\end{minipage}%
}%
\centering
\caption{The sum-rate and BER performances of the proposed scheme with respect to different numbers of users $K$ and antennas $N_{\rm BS}$ and different values of $L$.}
\vspace{-5mm}
\end{figure*}
We first present the sum-rate performance of the proposed scheme with respect to different numbers of antennas $N_{\rm BS}$ in Fig.~\ref{Sum_rate_vs_N_BS}, where we set $K=2$, $P=2$, and $L=1$.
As shown in the figure, the sum-rate increases by $K$ bits/s/Hz when doubling the number of antennas, which indicates a logarithmical increase of the sum-rate with with the number of antennas $N_{\rm BS}$ as indicated by Theorem~3.
The sum-rate performance for different numbers of users is presented in Fig.~\ref{Sum_rate_vs_K}, where we set $P=3$ and $L=1$. In particular, we apply a fixed ratio $\rho=2$ between the number of antennas $N_{\rm BS}$ and number of users $K$. It can be seen that the sum-rate appears to increase first with SNR and then slightly saturate in the very high SNR regime. This is because $L=1$ is not sufficient to perfectly cancel out the CTI for the considered case.
But we still observe that the sum-rate exhibits a strong increasing trend at practical SNRs, e.g., SNR from $10$ dB to $30$ dB. Furthermore, we also notice that with a fixed ratio $\rho$, the sum-rate is doubled if the number of users is doubled. This observation suggests a linear increase of the sum-rate with respect to the number of users $K$, and it is consistent with our findings in Theorem~2.

In Fig.~\ref{Sum_rate_vs_L}, the sum-rate performance with different values of $L$ is considered, where we set $N_{\rm BS}=8$, $K=4$, $P=3$. The performance bounds given in both~\eqref{MU_SY_sum_rate_finial} and~\eqref{Lemma2} are also drawn in the figure. As can be observed from the figure, the proposed scheme outperforms the no precoding benchmark in terms of the sum-rate. Furthermore, we also observe that the sum-rate increases with a larger $L$, but the rate saturation appears at very high SNRs. This is not unexpected because the number of CTI terms is large with a small antenna-to-user ratio and many resolvable paths. Consequently, a large $L$ is required to fully cancel the interference. On the other hand, it should be noticed that the sum-rate of the proposed scheme still shows a good increasing rate with imperfect cancellation at practical SNRs, e.g., SNR from $10$ dB to $30$ dB, as evidenced by the bounds. The choice of $L$ is important for the system designs, and more discussions on how to choose $L$ will be given later in Remark~1.

The bit error rate (BER) performance with various numbers of users, antennas, and resolvable paths is presented in Fig.~\ref{BER}, where we set $L=1$. As indicated by the figure, the BER performance with various channel conditions does not show a noticeable error floor at practical SNRs. Furthermore, we notice that increasing $P$ and $K$ could degrade the BER performance. This observation is consistent with the fact that more interference terms are introduced with an increasing number of resolvable paths and users. On the other hand, we also observe that the BER performance improves with an increasing number of BS antennas $N_{\rm BS}$. This observation is also consistent with our conclusions from Fig.~\ref{Sum_rate_vs_N_BS}.
\vspace{-3mm}
\subsection{Numerical Results under Practical Channel Conditions}
\begin{figure*}
\centering
\subfigure[Sum-rate performance for different $K$ and $N_{\rm BS}$.]{
\begin{minipage}[t]{0.5\textwidth}
\centering
\includegraphics[scale=0.5]{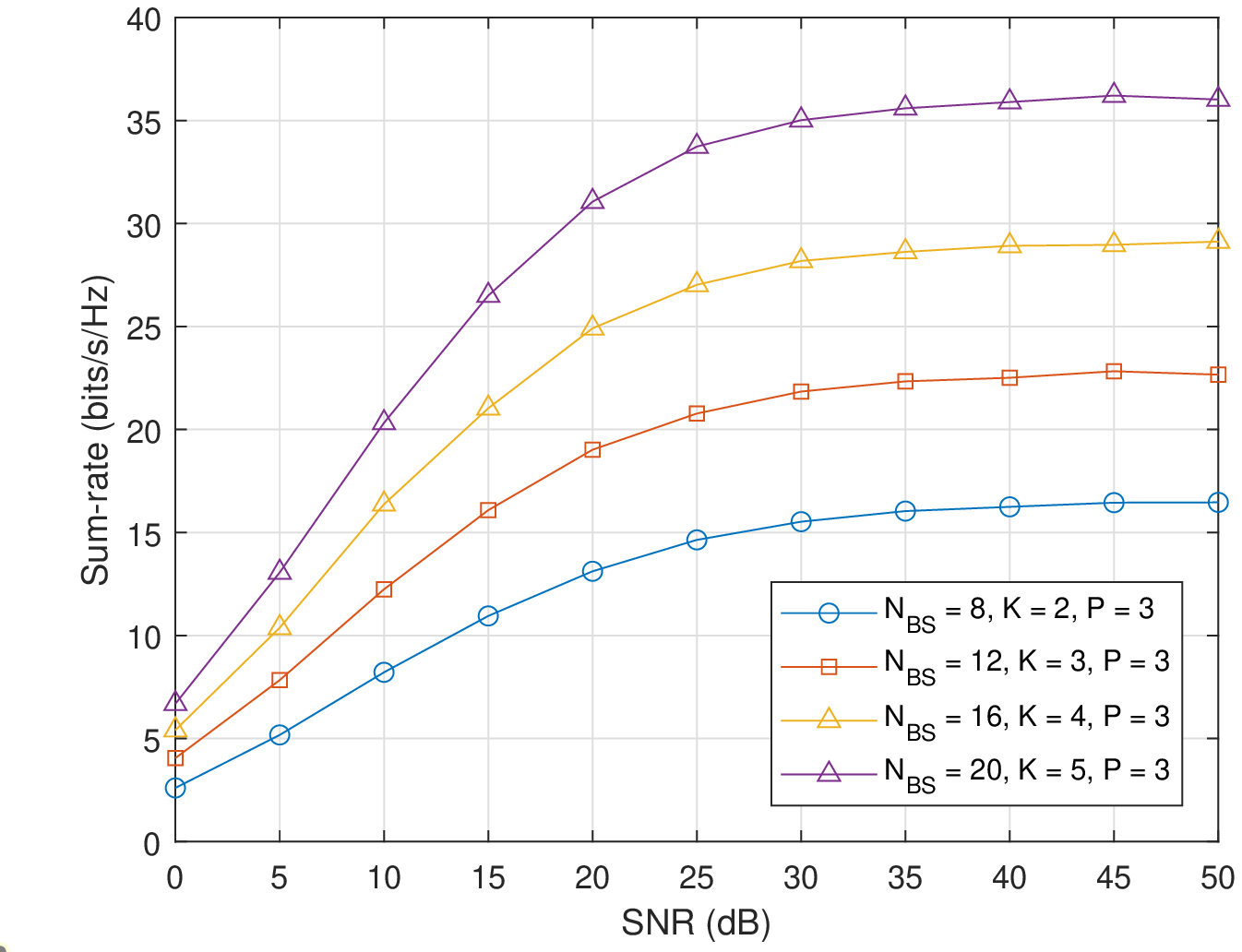}
\label{Sum_rate_vs_K_IBI_CTI}
\end{minipage}%
}%
\subfigure[Sum-rate comparison between various channel conditions.]{
\begin{minipage}[t]{0.5\textwidth}
\centering
\includegraphics[scale=0.5]{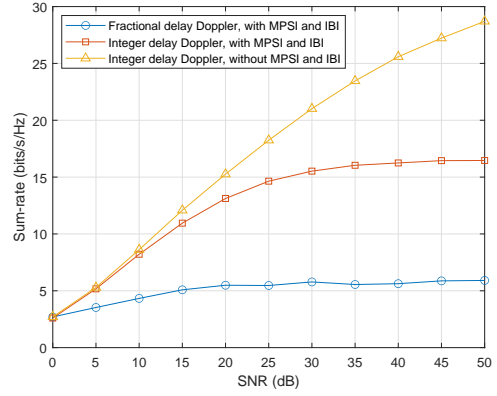}
\label{fractional_integer}
\end{minipage}%
}%
\quad
\subfigure[BER of THP, MRT~\cite{Pandey2021low}, and OFDM with ZF.]{
\begin{minipage}[t]{0.5\textwidth}
\centering
\includegraphics[scale=0.5]{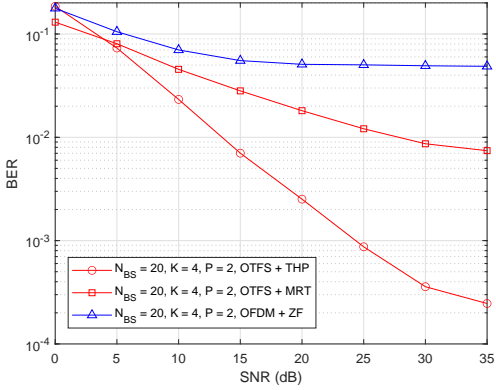}
\label{THP_MRT_OFDM_BER}
\end{minipage}%
}%
\subfigure[Sum-rates of THP, MRT~\cite{Pandey2021low}, and OFDM with ZF.]{
\begin{minipage}[t]{0.5\textwidth}
\centering
\includegraphics[scale=0.5]{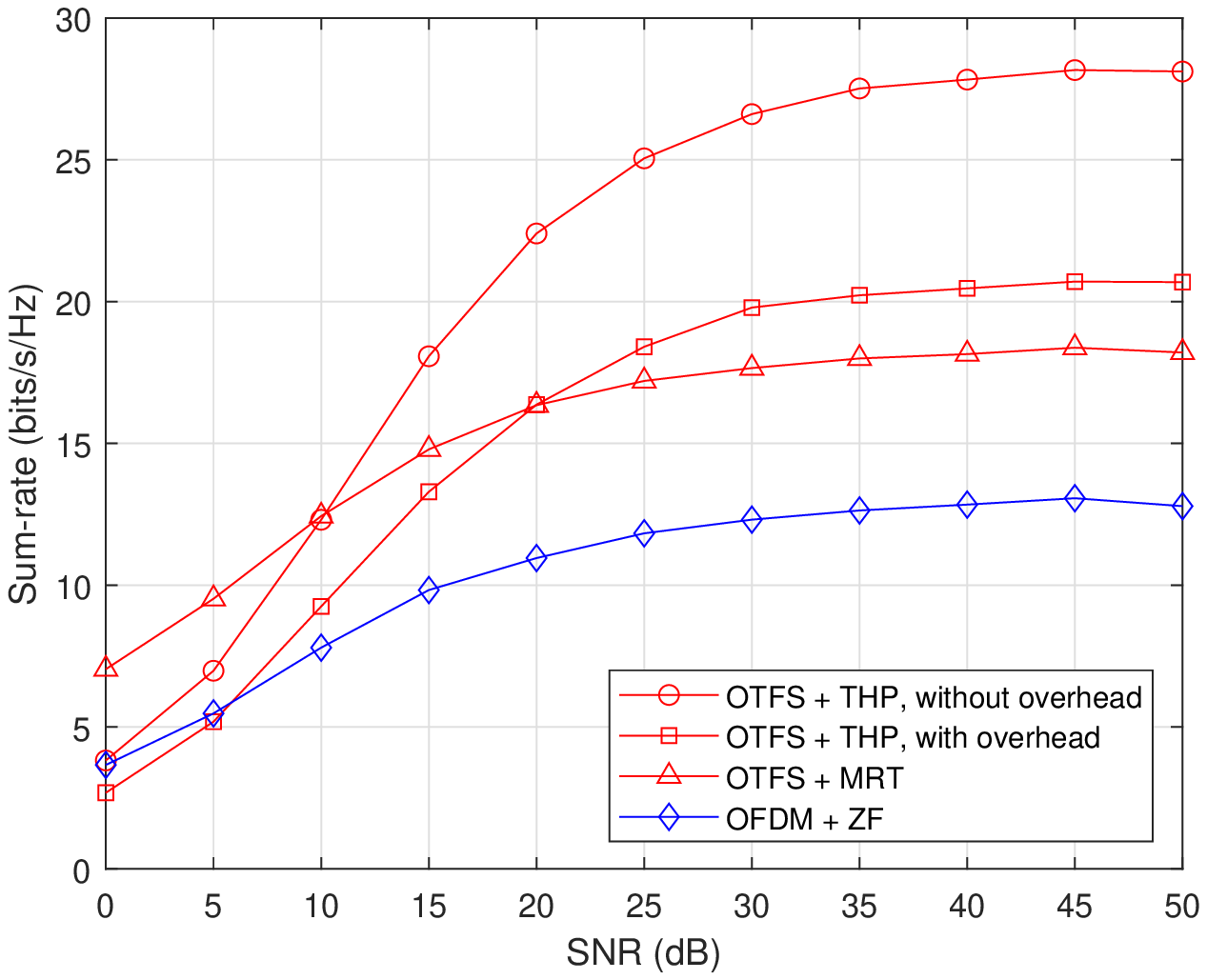}
\label{THP_MRT_OFDM_IBI_CTI}
\end{minipage}%
}%
\centering
\caption{The sum-rate performance of the proposed scheme with different parameters and benchmark technologies.}
\vspace{-5mm}
\end{figure*}
In this subsection, we present the numerical results of the proposed scheme under more realistic channel conditions, where both the MPSI and IBI are considered. We compare the sum-rate performance for different $K$ and $N_{\rm BS}$ in Fig.~\ref{Sum_rate_vs_K_IBI_CTI}, where $P=3$ and $L=1$. As can be observed from the figure, the sum-rate improves roughly linearly with the increase of $K$ at mid-to-high SNRs, but saturates when the SNR is larger than $30$ dB. This rate saturation is mainly caused by the MPSI and IBI. Note that the power of IBI and MPSI only relates to the transmitted signal power and the corresponding channel gain, and is independent from the noise power. Consequently, the rate saturation due to the interference will not be mitigated by a higher SNR, as reflected by the noticeable error-floor appeared at high SNRs.

We examine the proposed scheme with more complex channel conditions in Fig.~\ref{fractional_integer}, where we consider $N_{\rm BS}=8$, $K=3$, $P=4$, and $L=1$. In particular, we present the sum-rate performance with favorable propagation
(no MPSI and IBI), practical channel (with MPSI and IBI), and practical channel having fractional delay and Doppler. It can be observed that the proposed scheme enjoys a sum-rate increase with the growth of SNR even in the presence of fractional delay and Doppler. However, it suffers from a noticeable rate degradation, because the inter-Doppler and inter-delay interferences are treated as noise in the case of fractional delay and Doppler. It should be noted that the fractional delay and Doppler can be and should be dealt with by
baseband filtering, such as windowing~\cite{wei2021transmitter}, and pulse shaping~\cite{mohammed2021derivation,lampel2021orthogonal,Hai2022TWC,Wei2022CL,Shuangyang2022CL}.
On the other hand, we observe that the influence of MPSI and IBI becomes more severe at high SNRs, which aligns with the rate saturation observed from Fig.~\ref{Sum_rate_vs_K_IBI_CTI}.

A performance comparison between the proposed scheme, the MRT precoding in~\cite{Pandey2021low}, and OFDM with zero-forcing (ZF) precoding is presented in Fig.~\ref{THP_MRT_OFDM_BER} and Fig.~\ref{THP_MRT_OFDM_IBI_CTI}. To have a fair comparison, the OFDM also applies a reduced-CP structure, where no CP is appended between the adjacent OFDM symbols. But we apply a large ZF precoder of size $KN \times KN$ on each subcarrier to mitigate the intersymbol interference and multiuser interference. In Fig.~\ref{THP_MRT_OFDM_BER}, the BER performance of those schemes are presented, where we consider $N_{\rm BS}=20$, $K=4$, $P=2$, and $L=1$. It can be observed from the figure that the proposed scheme outperforms the MRT scheme and the OFDM with ZF at mid-to-high SNRs. This observation validates the advantage of the proposed THP over existing schemes.
This advantage can also be demonstrated by the achieved sum-rate gain shown in Fig.~\ref{THP_MRT_OFDM_IBI_CTI}, where we consider $N_{\rm BS}=8$, $K=4$, $P=3$, and $L=1$.
In particular, we also include the sum-rate results of the proposed THP with and without considering the required overhead in Fig.~\ref{THP_MRT_OFDM_IBI_CTI}.
We observe from the figure that the proposed scheme exhibits roughly the same sum-rate as the ZF precoded OFDM at relatively low SNRs, which is lower than the MRT precoded OTFS. Note that the achievable rate of THP generally suffers from the ``modulo loss'' at low SNRs, which is due to the modulo operation applied at the receiver~\cite{Wesel1998Achievable}.
However, this rate loss decreases quickly with an increased SNR as shown in the figure.
We notice that, at high SNRs, the proposed scheme outperforms the existing schemes in terms of the sum-rate, even when the overhead is considered. It should be highlighted that the required overhead can be reduced as discussed in Section III-D, which is a topic for future research.
More importantly, the proposed THP only requires a linear complexity of ${\cal O}\left( LKMN \right)$, while the MRT in~\cite{Pandey2021low} requires matrix/vector superposition and multiplication, thus having a complexity of ${\cal O}\left( KM^2N^2 \right)$. Furthermore, the ZF precoded OFDM requires the matrix inversion and has a complexity of ${\cal O}\left( MK^3N^3\right)$. The superior performance and the low implementation complexity make our proposed THP a promising candidate for downlink MU-MIMO transmissions.

\textbf{Remark~1}: The pre-cancellation term $L$ is a key parameter for our proposed THP, which determines how many CTI interference terms are pre-cancelled in the precoding. Note that the value of $L$ should be selected considering the channel condition, operating SNR, and the cancellation strategy discussed in Section III-D. In our simulations, we intentionally use small values of $L$, such as $L=1$, because this is the most straightforward application of the proposed THP and it also requires the least overhead. As extensively discussed in our numerical results, $L = 1$ performs quite well under various channel conditions. We argue that this is not a coincidence. Instead, this is an expected result due to the careful user grouping strategy. The important insight here is that the CTI interference is only severe when the BF path of one user has a direction that is sufficiently close to the non-BF path of a different user, as depicted in Fig.~\ref{THP_interference_model}. Therefore, it is almost impossible that the BF paths of different users have similar AoDs overlapping with the same non-BF path of a specific user after a reasonable user grouping. Furthermore, the possibility of multiple users' BF paths overlapping with different non-BF paths of the same user is generally low, and this case can also be avoided by smart grouping strategy. Therefore, we can safely choose a relatively small value of $L$ in practical systems facilitated by a carefully grouping of users.

\begin{table}[]
\caption{Overhead vs. different numbers of users and resolvable paths.}
\centering
\begin{tabular}{|c|c|c|c|}
\hline
      & $K=2, L=1$ & $K=3, L=1$ & $K=3, L=2$ \\ \hline
$P=2$ & $2.9\%$    & $24.1\%$   & $34.9\%$   \\ \hline
$P=3$ & $9.6\%$    & $25.2\%$   & $37.8\%$   \\ \hline
$P=4$ & $12.9\%$   & $25.7\%$   & $39.0\%$   \\ \hline
\end{tabular}
\label{overhead}
\vspace{-3mm}
\end{table}

\textbf{Remark~2}: It is important to evaluate the required overhead of the proposed scheme.
In Table~\ref{overhead}, we compute the overhead of the proposed scheme with $N_{\rm BS}=16$ and different $K$ and $L$. The overhead is calculated as the ratio between the number of assigned known symbols in the DD domain and the number of DD grids in total, i.e., $KMN$, which is represented in the form of a percentage. We observe that the overhead generally increases with more resolvable paths and users, due to the increase of interference terms. On the other hand, we also notice that a larger value of $L$ also increases the overhead. However,
we have discussed in Remark~1 that a relatively small value of $L$ is sufficient in practical systems, which is also consistent with our numerical results in this section.
Furthermore, it should be noted that the overhead performance can be further improved by considering the scheduling of pre-cancellation as discussed in Section III-D.
\section{Conclusions}
In this paper, we investigated the DD domain THP for MU-MIMO-OTFS. In particular, the proposed THP implementation exploits the DD domain channel characteristics and does not require any matrix decomposition or inversion. Furthermore, we analyzed performance for the proposed scheme in terms of the achievable rates and investigated the scaling factors for the number of BS antennas and users. Our derivations implied that the sum-rate increases logarithmically with the number of antennas and linearly with the number of users (under the same antenna-to-user ratio). Our derivations were verified by numerical results. Our future work may investigate overhead reduction approaches for DD domain THP.
\appendices

\section*{Acknowledgement}
The authors would like to express their thanks to the inventor of OTFS modulation, Prof. Ronny Hadani, for his enlightening speech on MU-MIMO-OTFS, which motivates this work.

\bibliographystyle{IEEEtran}
\bibliography{OTFS_references}

\end{document}